\documentclass[aps,twocolumn]{revtex4}% Physical Review B
\usepackage{bm}
\usepackage{epsfig}

\newcommand{\nix}[1]{}

\def\dfrac{\displaystyle\frac}

\begin{document}

\title{Magneto-Gyrotropic Photogalvanic Effects in Semiconductor Quantum Wells }
\author{V.V.~Bel'kov$^{1,2}$, S.D.~Ganichev$^{1,2}$,  E.L.~Ivchenko$^2$,
S.A.~Tarasenko$^2$,  W.~Weber$^{1}$, S.~Giglberger$^1$,
M.~Olteanu$^1$, H.-P.~Tranitz$^1$, S.N.~Danilov$^1$,
Petra~Schneider$^1$, W.~Wegscheider$^1$, D.~Weiss$^1$, and
W.~Prettl$^1$}
\affiliation{\vskip 4pt $^1$Fakult\"{a}t Physik, University of
Regensburg, 93040, Regensburg, Germany,}
\affiliation{$^2$A.F.~Ioffe Physico-Technical Institute, Russian
Academy of Sciences, 194021 St.~Petersburg, Russia}

%\date{\today}

\begin{abstract}
We show that free-carrier (Drude) absorption of both polarized and
unpolarized terahertz radiation in quantum well (QW) structures
causes an electric photocurrent in the presence of an in-plane
magnetic field. Experimental and theoretical  analysis evidences
that the observed photocurrents are spin-dependent and related to
the gyrotropy of the QWs. Microscopic models for the photogalvanic
effects in QWs based on asymmetry of photoexcitation and
relaxation processes are proposed. In most of the investigated
structures the observed magneto-induced photocurrents are caused
by spin-dependent relaxation of non-equilibrium carriers.
\end{abstract}
\pacs{73.21.Fg, 72.25.Fe, 78.67.De, 73.63.Hs}

\maketitle

\tableofcontents

%%%%%%%%%%%%%%%%%%%%%%%%%%%%%%%%%%%%%%%%%%%%%%%%%%%%%%%%%%%%%%%%%
\section{Introduction}

Much current interest in condensed matter physics is directed
towards understanding of spin dependent phenomena. In particular,
the spin of electrons and holes in solid state systems is the
decisive ingredient for spintronic devices~\cite{Wolf}. Recently
spin photocurrents generated in QWs and bulk materials have
attracted considerable attention~\cite{ufn,review2003spin}. Among
them are currents caused by a gradient of a spin-polarized
electron
density~\cite{Averkiev83p393,Dyakonov71p144,Bakun84p1293}, the
spin-galvanic effect~\cite{Nature02}, the circular photogalvanic
effect in QWs~\cite{PRL01}, pure spin currents under simultaneous
one- and two-photon coherent
excitation~\cite{Bhat00p5432,Stevens02p4382} and spin-polarized
currents due to the photo-voltaic effect in $p$-$n$
junctions~\cite{Zutic01p1558}. Experimentally, a natural way to
generate spin photocurrents is the optical excitation with
circularly polarized radiation. The absorption of circularly
polarized light results in optical spin orientation of free
carriers due to a transfer of photon angular momenta to the
carriers~\cite{Meier}. Because of the spin-orbit coupling such
excitation may result in an electric current. A characteristic
feature of this electric current is that it reverses its direction
upon changing the radiation helicity from left-handed to
right-handed and vice versa.

However, in an external magnetic field spin photocurrents may be
generated even by unpolarized radiation as it has been proposed
for bulk gyrotropic crystals~\cite{cpge,bulli}. Here we report on
an observation of these spin photocurrents in QW structures caused
by the Drude absorption of terahertz radiation. We show that,
microscopically, the effects under study are related to the
gyrotropic properties of the structures. The gyrotropic point
group symmetry makes no difference between components of axial and
polar vectors, and hence allows an electric current $j \propto I
B$, where $I$ is the light intensity and $B$ is the applied
magnetic field. Photocurrents which require simultaneously
gyrotropy and the presence of a magnetic field may be gathered in
a class of magneto-optical phenomena denoted  as
magneto-gyrotropic photogalvanic effects. So far such currents
were intensively studied in low-dimensional structures at direct
inter-band and inter-subband
transitions~\cite{magarill,gorbats,kibis,spivak,emelya,emelya1,moscow,kucher}.
In these investigations the magneto-induced photocurrents were
related to spin independent mechanisms, except for
Refs.~\cite{magarill,emelya1} where direct optical transitions
between branches of the spin-split electron subband were
considered. This mechanism requires, however, the spin splitting
and the photon energy to be comparable whereas, in the conditions
under study here, the spin splitting is much smaller than the
photon energy and the light absorption occurs due to indirect
(Drude-like) optical transitions. It is clear that
magneto-gyrotropic effects due to the Drude absorption may also be
observed at excitation in the microwave range
 where the basic mechanism is free carrier absorption as well.
This could link electronics to spin-optics. In most of the
investigated structures, the photogalvanic measurements reveal a
magneto-induced current which is independent of the direction of
light in-plane linear polarization and related to spin-dependent
relaxation of non-equilibrium carriers. In addition, our results
show that, without a magnetic field, non-equilibrium free carrier
heating can be accompanied by spin flow similar to spin currents
induced in experiments with simultaneous one- and two-photon
coherent excitation~\cite{Stevens02p4382} or in the spin Hall
effect~\cite{Wunderlicht2004,Kato2004}.

\section{Phenomenological theory}

Illumination of gyrotropic nanostructures in the presence of a
magnetic field may result in a photocurrent. There is a number of
contributions to the magnetic field induced photogalvanic effect
whose microscopic origins will be considered in
Section~\ref{micmodel}. The contributions are characterized by
different dependencies of the photocurrent magnitude and direction
on the radiation polarization state and the orientation of the
magnetic field with respect to the crystallographic axes. As a
consequence, a proper choice of experimental geometry allows to
investigate each contribution separately. Generally, the
dependence of the photocurrent on the light polarization and
orientation of the magnetic field may be obtained from
phenomenological theory which does not require knowledge of the
microscopic origin of the current. Within the linear approximation
in the magnetic field strength $\bm{B}$, the magneto-photogalvanic
effect (MPGE) is given by
\begin{equation} \label{phen0}
j_\alpha = \sum_{\beta\gamma\delta}
\phi_{\alpha\beta\gamma\delta}\:B_\beta\:\{E_\gamma
E^\star_\delta\} + \sum_{\beta\gamma}
\mu_{\alpha\beta\gamma}\:B_\beta
\hat{e}_\gamma\:E^2_0\:P_{circ}\:.
\end{equation}
Here the fourth rank pseudo-tensor $\bm{\phi}$ is symmetric in the
last two indices, $E_\gamma$ are components of the complex
amplitude of the radiation electric field $\bm E$. In the
following the field is presented as $E = E_0 \bm{e} $ with $E_0$
being the modulus $|\bm{E}|$ and $\bm{e}$ indicating the (complex)
polarization unit vector, $|\bm{e}| = 1$. The symbol $\{E_\gamma
E^\star_\delta\}$ means the symmetrized product of the electric
field with its complex conjugate,
\begin{equation}\label{sym}
\{E_\gamma  E^\star_\delta\}  = \frac{1}{2}\left(E_\gamma
E^\star_\delta + E_\delta  E^\star_\gamma\right) \:.
\end{equation}
In the second term on the right hand side of Eq.~(\ref{phen0}),
$\bm{\mu}$ is a regular third rank tensor, $P_{circ}$ is the
helicity of the radiation and $\bm{\hat e}$ is the unit vector
pointing in the direction of light propagation. While the second
term requires circularly polarized radiation the first term may be
non-zero even for unpolarized radiation.

We consider (001)-oriented QWs based on zinc-blende-lattice
III-V or II-VI compounds. Depending on the equivalence or
non-equivalence of the QW interfaces their symmetry may belong to
one of the point groups $D_{2d}$ or $C_{2v}$, respectively. The
present experiments have been carried out on the $C_{2v}$ symmetry
structures and, therefore, here we will focus on them only.

For the $C_{2v}$ point group, it is convenient to write the
components of the magneto-photocurrent in the coordinate system
with $x^\prime \parallel [1 \bar{1} 0]$ and $y^\prime \parallel
[110]$
%
%and $z\parallel [001]$
%
or in the system $x \parallel [1 0 0]$ and $y
\parallel [010]$. The advantage of the former system is
that the in-plane axes $x', y'$ lie in the crystallographic planes
$(110)$ and $(1 \bar{1} 0)$ which are the mirror reflection planes
containing the two-fold axis $C_2$. In the system $x^\prime,
y^\prime, z$ for normal incidence of the light and the in-plane
magnetic field, Eq.~(\ref{phen0}) is reduced to
\begin{widetext}
\begin{eqnarray}
\label{phen} j_{x^\prime} = S_1  B_{y^\prime}I + S_2 B_{y^\prime}
\left( |e_{x^\prime}|^2 - |e_{y^\prime}|^2 \right) I+ S_3
B_{x^\prime} \left( e_{x^\prime} e^*_{y^\prime} + e_{y^\prime}
e^*_{x^\prime} \right)
I+ S_4 B_{x^\prime}   I P_{\rm circ}\:,\\
j_{y^\prime} = S'_1  B_{x^\prime}I + S'_2 B_{x^\prime}  \left(
|e_{x^\prime}|^2 - |e_{y^\prime}|^2 \right) I+ S'_3 B_{y^\prime}
\left( e_{x^\prime} e^*_{y^\prime} + e_{y^\prime} e^*_{x^\prime}
\right) I + S'_4  B_{y^\prime}I P_{\rm circ} \:, \nonumber
\end{eqnarray}
\end{widetext}
where, for simplicity, we set for the intensity $I=E_0^2$. The
parameters $S_1$ to $S_4$ and $S^\prime_1$ to $S^\prime_4$
expressed in terms of non-zero components of the tensors
$\bm{\phi}$ and $\bm{\mu}$ allowed by the $C_{2v}$ point group are
given in Table~\ref{t1}. The first terms on the right hand side of
Eqs.~(\ref{phen}) (described by $S_1, S'_1$) yield a current in
the QW plane which is independent of the radiation polarization.
This current is induced even by unpolarized radiation. Each
following contribution has a special polarization dependence which
permits to separate it experimentally from the others.

%%%%%TABLE%%%%%%%%%%%%%%%%%%%%%%%%%%%%%%%%%%%%%%%%%%%%%%%%%%%%%%%%%%%%%
\begin{table}[t]
\renewcommand{\arraystretch}{1.5}
\begin{tabular}{|r@{=}l|r@{=}l|}
\hline

$S_1$ &
$\frac{1}{2}(\phi_{{x^\prime}{y^\prime}{x^\prime}{x^\prime}}+
\phi_{{x^\prime}{y^\prime}{y^\prime}{y^\prime}})$ & $S^\prime_1$ &
$\frac{1}{2}(\phi_{{y^\prime}{x^\prime}{x^\prime}{x^\prime}}+
\phi_{{y^\prime}{x^\prime}{y^\prime}{y^\prime}})$ \\\hline

$S_2$ &
$\frac{1}{2}(\phi_{{x^\prime}{y^\prime}{x^\prime}{x^\prime}}-
\phi_{{x^\prime}{y^\prime}{y^\prime}{y^\prime}})$ & $S^\prime_2$ &
$\frac{1}{2}(\phi_{{y^\prime}{x^\prime}{x^\prime}{x^\prime}}-
\phi_{{y^\prime}{x^\prime}{y^\prime}{y^\prime}})$
\\\hline

$S_3$ &
$\phi_{{x^\prime}{x^\prime}{x^\prime}{y^\prime}}=\phi_{{x^\prime}{x^\prime}{y^\prime}{x^\prime}}$
& $S^\prime_3$ &
$\phi_{{y^\prime}{y^\prime}{x^\prime}{y^\prime}}=\phi_{{y^\prime}{y^\prime}{y^\prime}{x^\prime}}$\\\hline

$S_4$ & $\mu_{{x^\prime}{x^\prime}z}$ & $S^\prime_4$ &
$\mu_{{y^\prime}{y^\prime}z}$
\\\hline
\end{tabular}
\caption{Definition of the parameters $S_i$ and $S^\prime_i$
($i=1\dots4$) in Eqs.~(\ref{phen}) in terms of non-zero components
of the tensors $\bm{\phi}$ and $\bm{\mu}$ for the coordinates
${x^\prime}
\parallel [1 \bar{1} 0]$, ${y^\prime} \parallel [110]$ and $z\parallel
[001]$. The $C_{2v}$ symmetry and normal incidence of radiation
along $z$ are assumed.} \label{t1}
\end{table}
%%%%%%%%%%%%%%%%%%%%%%%%%%%%%%%%%%%%%%%%%%%%%%%%%%%%%%%%%%%%%%%%%%%%%%%%%%%%

%%%%%TABLE%%%%%%%%%%%%%%%%%%%%%%%%%%%%%%%%%%%%%%%%%%%%%%%%%%%%%%%%
\begin{table}
\renewcommand{\arraystretch}{1.5}
\begin{tabular}{|r@{=}l|r@{=}l|}
\hline

$S^+_1$ & $\frac{1}{2}(\phi_{xxxx} +\phi_{xxyy})$ & $S^-_1$ & $\frac{1}{2}(\phi_{xyxx} +\phi_{xyyy})$ \\

& $-\frac{1}{2}(\phi_{yyxx} +\phi_{yyyy})$ & &
$-\frac{1}{2}(\phi_{yxxx} +\phi_{yxyy})$ \\\hline

$S^+_2$ & $\phi_{yyxy} = \phi_{yyyx} $&  $S^-_2$ & $\phi_{yxxy} = \phi_{yxyx}$ \\

& $-\phi_{xxxy} =- \phi_{xxyx} $&  & $-\phi_{xyxy} = -\phi_{xyyx}
$\\\hline

$S^+_3$ & $\frac{1}{2}(\phi_{xxxx} -\phi_{xxyy})$ & $S^-_3$ & $-\frac{1}{2}(\phi_{xyxx} -\phi_{xyyy})$ \\

& $\frac{1}{2}(\phi_{yyxx} - \phi_{yyyy})$ & &
$-\frac{1}{2}(\phi_{yxxx} -\phi_{yxyy})$ \\\hline

$S^+_4$ & $\mu_{xxz} = \mu_{yyz}$ & $S^-_4$ & $-\mu_{xyz} =
-\mu_{yxz}$\\\hline
\end{tabular}
\caption{Definition of the parameters $S^+_i$ and $S^-_i$
($i=1\dots4$) in Eqs.~(\ref{phena}) in terms of non-zero
components of the tensors $\bm{\phi}$ and $\bm{\mu}$ for the
coordinates $x\parallel [100]$, $y\parallel [010]$ and $z\parallel
[001]$. The $C_{2v}$ symmetry and normal incidence of radiation
along $z$ are assumed. } \label{t1a}
\end{table}
%%%%%%%%%%%%%%%%%%%%%%%%%%%%%%%%%%%%%%%%%%%%%%%%%%%%%%%%%%%%%%%%%%%%%%%%%%%%%%%%%%%%%

\paragraph*{ Linearly polarized radiation.}
For {\em linearly} polarized light, the terms described by
parameters $S_2, S'_2$ and $S_3, S'_3$ are proportional to
$|e_{x^\prime}|^2 - |e_{y^\prime}|^2 = {\rm cos}\,2\alpha$ and $
e_{x^\prime} e^*_{y^\prime} + e_{y^\prime} e^*_{x^\prime} = {\rm
sin}\,2\alpha$, respectively, where $\alpha$ is the angle between
the plane of linear polarization and the $x'$ axis. Hence the
current reaches maximum values for light polarized either along
${x^\prime}$ or ${y^\prime}$ for the second terms (parameters
$S_2, S'_2$), or along the bisector of ${x^\prime}$, ${y^\prime}$
for the third terms, proportional to $S_3, S'_3$. The last terms
(parameters $S_4, S'_4$),  being proportional to $P_{circ}$,
vanish for {\em linearly} polarized excitation.

\paragraph*{Elliptically polarized radiation.}

For {\em elliptically} polarized light all contributions are
allowed. In the experiments discussed below, elliptically and, in
particular, circularly polarized radiation was achieved by passing
laser radiation, initially linearly polarized along ${x^\prime}$
axis, through a $\lambda/4$-plate. Rotation of the plate results
in a variation of both linear polarization and helicity as follows
\begin{equation}
\label{plin} P_{\rm lin} \equiv \frac12 (e_{x^\prime}
e^*_{y^\prime} + e_{y^\prime} e^*_{x^\prime}) = \frac14 \sin{4
\varphi}\:,\:
\end{equation}
\vspace{-0.7cm}
\begin{equation} \label{plinprime}
P'_{\rm lin} \equiv \frac12 (|e_{x^\prime}|^2 - |e_{y^\prime}|^2)
= \frac{1 + \cos{4 \varphi} }{4}\:,\:
\end{equation}
\vspace{-0.7cm}
\begin{equation}
\label{circ}
 P_{\rm circ} = \sin{2 \varphi} \:.
\end{equation}
Two Stokes parameters $P_{\rm lin}, P'_{\rm lin}$ describe the
degrees of linear polarization and $\varphi$ is the angle between
the optical axis of $\lambda/4$ plate and the direction of the
initial polarization ${x^\prime}$.

As described above, the first terms on the right hand side of
Eqs.~(\ref{phen}) are independent of the radiation polarization.
The polarization dependencies of magneto-induced photocurrents
caused by  second and third terms in Eqs.~(\ref{phen}) are
proportional to $P'_{\rm lin}$ and $P_{\rm lin}$, respectively.
These terms vanish if the radiation is circularly polarized. In
contrast, the last terms in Eqs.~(\ref{phen}) describe a
photocurrent proportional to the helicity of radiation. It is zero
for linearly polarized radiation and reaches its maximum for
 left- or right-handed circular polarization. Switching
helicity $P_{circ}$ from $+1$ to $-1$ reverses the current
direction.

As we will see below  the photocurrent analysis for $x
\parallel [100]$ and $y \parallel [010]$  directions helps to conclude
on the microscopic nature of the different contributions to the
MPGE. In these axes Eqs.~(\ref{phen}) read
\begin{widetext}
\begin{equation}\label{phena}
j_x = S_1^+ B_x I + S_1^- B_y I - (S_2^+ B_x + S_2^- B_y) \left(
e_x e_y^* + e_y e_x^* \right) I + ( S_3^+ B_x - S_3^- B_y ) \left(
|e_x|^2 - |e_y|^2 \right) I  + ( S_4^+ B_x - S_4^- B_y ) I P_{\rm
circ}\:,
\end{equation}
\vspace{-0.8cm}
\[
j_y = - S_1^- B_x I - S_1^+ B_y I + (S_2^- B_x + S_2^+ B_y) \left(
e_x e_y^* + e_y e_x^* \right)I + ( - S_3^- B_x + S_3^+ B_y )
\left( |e_x|^2 - |e_y|^2 \right) I  + ( - S_4^- B_x + S_4^+ B_y )
I P_{\rm circ}\:,
\]
\end{widetext}
where $S_l^{\pm} = (S_l \pm S'_l)/2$ ($l=1\dots4$). The parameters
$S^{\pm}_1$ to $S^{\pm}_4$ expressed via non-zero elements of the
tensors $\bm{\phi}$ and $\bm{\mu}$ for the $C_{2v}$ symmetry are
given in Table~\ref{t1a}. Equations~(\ref{phena}) show that, for a
magnetic field oriented along a cubic axis, all eight parameters
$S_l^{\pm}$ contribute to the photocurrent components, either
normal or parallel to the magnetic field. However, as well as for
the magnetic field oriented along $x^\prime$ or $y^\prime$ the
partial contributions can be separated analyzing polarization
dependencies.

For the sake of completeness, in Appendices A and B we present the
phenomenological equations for the magneto-photocurrents in the
systems of the T$_d$ and C$_{\infty v}$ symmetries, respectively, representing
the bulk zinc-blende-lattice semiconductors and
axially-symmetric QWs with nonequivalent interfaces.

Summarizing the macroscopic picture we note that, for normal
incidence of the radiation on a (001)-grown QW, a magnetic field
applied in the interface plane is required to obtain a
photocurrent. In Table~\ref{t2} we present the relations between
the photocurrent direction, the state of light polarization and
the magnetic field orientation which follow from Eqs.~(\ref{phen})
and Eqs.~(\ref{phena}) and determine the appropriate experimental
geometries (Section~\ref{experim}). In order to ease data analysis
we give in Table~\ref{t5} polarization dependencies for geometries
relevant to experiment. Specific polarization behavior of each
term allows to discriminate between different terms in
Eqs.~(\ref{phen}).

\begin{widetext}
%
%%%%%TABLE%%%%%%%%%%%%%%%%%%
\begin{center}
\begin{table}[t]

        \begin{tabular}{|c|c|c|c|c|c|}
        \hline
        \multicolumn{2}{|l|}{\raisebox{3ex}[-3ex] {} } & \mbox{\bf 1$^{st}$ term} & \mbox{\bf 2$^{nd}$ term} & \mbox{\bf 3$^{rd}$ term} & \mbox{\bf 4$^{th}$ term} \\\hline\hline

                \rule[-3mm]{0mm}{9mm}

        & $j_{x^\prime}/I$ & 0 & 0 &$S_3 B_{x^\prime} \left(e_{x^\prime}e^*_{y^\prime}+e_{y^\prime}e^*_{x^\prime}\right)$ & $S_4 B_{x^\prime} P_{\rm{circ}}$ \\ \cline{2-6}
        \rule[-3mm]{0mm}{9mm}
        \raisebox{3ex}[-3ex] {$B\|x^\prime$} & $j_{y^\prime}/I$ & $S^\prime_1 B_{x^\prime}$ & $S^\prime_2 B_{x^\prime} \left(|e_{x^\prime}|^2-|e_{y^\prime}|^2\right)$ & 0 & 0 \\ \hline\hline

                \rule[-3mm]{0mm}{9mm}
        & $j_{x^\prime}/I $&$ S_1 B_{y^\prime}$ & $S_2 B_{y^\prime} \left(|e_{x^\prime}|^2-|e_{y^\prime}|^2\right)$  & 0 & 0 \\ \cline{2-6}
        \rule[-3mm]{0mm}{9mm}
        \raisebox{3ex}[-3ex] {$B\|{y^\prime}$} & $j_{y^\prime}/I $& 0 & 0 & $S^\prime_3 B_{y^\prime} \left(e_{x^\prime}e^*_{y^\prime}+e_{y^\prime}e^*_{x^\prime}\right) $ & $S^\prime_4 B_{y^\prime} P_{\rm{circ}} $  \\ \hline\hline

                \rule[-3mm]{0mm}{9mm}
        & $j_{x}/I$ & $S^+_1 B_x$ & $-S^+_2B_x \left(e_xe^*_y+e_ye^*_x\right) $& $S^+_3B_x \left(|e_x|^2-|e_y|^2\right)$ & $S^+_4 B_x  P_{\rm{circ}}$ \\ \cline{2-6}
        \rule[-3mm]{0mm}{9mm}
        \raisebox{3ex}[-3ex] {$B\|x$} &$ j_{y}/I $& $-S^-_1B_x$ &$ S^-_2B_x \left(e_x e^*_y+e_y e^*_x\right)$ & $-S^-_3B_x \left(|e_x|^2-|e_y|^2\right)$ & $-S^-_4 B_x P_{\rm{circ}} $ \\ \hline\hline

                \rule[-3mm]{0mm}{9mm}
        & $j_{x}/I$ & $S^-_1B_y$ & $-S^-_2B_y \left(e_xe^*_y+e_ye^*_x\right)$ & $-S^-_3B_y \left(|e_x|^2-|e_y|^2\right)$ &$ -S^-_4 B_y P_{\rm{circ}}$  \\ \cline{2-6}
        \rule[-3mm]{0mm}{9mm}
        \raisebox{3ex}[-3ex] {$B\|y$} & $j_{y}/I$ &$ -S^+_1B_y $& $S^+_2B_y \left(e_xe^*_y+e_ye^*_x\right)$ & $S^+_3B_y \left(|e_x|^2-|e_y|^2\right)$ & $S^+_4 B_y P_{\rm{circ}}$  \\ \hline

        \end{tabular}
\caption{Contribution of the different terms in Eqs.~(\ref{phen})
and Eqs.~(\ref{phena}) to the current at different magnetic field
orientations. The two left columns indicate the magnetic field
orientation and the photocurrent component, respectively.}
\label{t2}
\end{table}
\end{center}
%%%%%%%%%%%%%%%%%%%%%%%%%%%%%%%%%%%%%%%%%%%%%%%%%%%%%%%%%%%%%%%%%%%%%%%%%%%%%%
%
\end{widetext}

%%%%%%%%TABLE%%%%%%%%%%%%%%%%%%%%%%%%%%%%%%
%\begin{center}
\begin{table}
        \begin{tabular}{|c|c|c|c|c|}
        \hline
         % \multicolumn{2}{|l|}{\raisebox{3ex}[-3ex] {} }
         & \mbox{\bf 1$^{st}\,$term} & \mbox{\bf 2$^{nd}$ term} & \mbox{\bf 3$^{rd}$ term} & \mbox{\bf 4$^{th}$ term} \\\hline\hline
        \rule[-3mm]{0mm}{9mm}
        $j_{x^\prime}(\varphi)$& $S_1 B_{y^\prime}$ & $S_2 B_{y^\prime}\dfrac{1 + {\rm cos}\,4\varphi}{2}$ & 0 & 0 \\ \hline %\cline{2-6}
        \rule[-3mm]{0mm}{9mm}
     %   \raisebox{3ex}[-3ex] {$j\|{x^\prime}$}
        $j_{x^\prime}(\alpha) $& $S_1 B_{y^\prime}$ & $S_2 B_{y^\prime}\,{\rm cos}\,2\alpha$ & 0 & 0  \\ \hline \hline
                \rule[-3mm]{0mm}{9mm}
        $j_{y^\prime}(\varphi)$ & 0 & 0 &$S^\prime_3B_{y^\prime}\, \dfrac{{\rm sin}\,4\varphi}{2}$ & $S^\prime_4 B_{y^\prime}\,{\rm sin}\,2\varphi$ \\ \hline %\cline{2-6}
        \rule[-3mm]{0mm}{9mm}
      %  \raisebox{3ex}[-3ex] {$j\|y^\prime$}
        $j_{y^\prime}(\alpha)$ & 0 & 0 & $S^\prime_3B_{y^\prime}\,{\rm sin}\,2\alpha$ & 0 \\ \hline %\hline
        \end{tabular}
\caption{Polarization dependencies of different terms in
Eqs.~(\ref{phen}) at $\bm{B}\, \| \, y'$. } \label{t5}
\end{table}
%\end{center}
%%%%%%%%%%%%%%%%%%%%%%%%%%%%%%%%%%%%%%%%%%%%%%%%%%%%%%%%%%%%%%%%%%%%%%%%%%%%%%%%%%%%
%

\section{Methods}

The experiments were carried out on MBE-grown (001)-oriented
$n$-type GaAs/Al$_{0.3}$Ga$_{0.7}$As and InAs/AlGaSb QW
structures. The characteristics of the investigated samples are
given in Table~\ref{tsample}. The InAs/AlGaSb heterostructure were
grown on a semi-insulating GaAs substrate. The quantum well is
nominally undoped, but contains a two dimensional electron gas
with the carrier density of $8\cdot 10^{11}$~cm$^{-2}$ at 4.2~K
located in the InAs channel. Details of the growth procedure are
given in~\cite{Beneth98p426}. All GaAs samples are
modulation-doped. For samples A2$-$A4 Si-$\delta$-doping, either
one-sided with spacer layer thicknesses of 70~nm (A3) and 80~nm
(A4), or double-sided with 70~nm spacer layer thickness (A2), has
been used. In contrast, for sample A5 the AlGaAs barrier layer
separating the QWs has been homogeneously Si-doped on a length of
30 nm. In the sample with a QW separation of 40 nm, this results
in a spacer thickness of only 5 nm. Therefore, in addition to the
different impurity distribution compared to the samples A2$-$A4,
the sample A5 has much lower mobility.

All samples have two pairs of ohmic contacts at the corners
corresponding to the $x \parallel [100]$ and $y
\parallel [010]$ directions, and two additional pairs of contacts centered
at opposite sample edges with the connecting lines along $x^\prime
\parallel [1\bar{1}0]$ and $y^\prime \parallel [110]$
(see inset in Fig.~\ref{fig1f}). The external magnetic field $B$
up to $1$\,T was applied parallel to the interface plane.

A pulsed optically pumped terahertz laser was used for optical
excitation~\cite{PhysicaB99tun}. With  NH$_3$ as active gas 100~ns
pulses of linearly polarized radiation with $\sim$10~kW power have
been obtained at  wavelengths 148~$\mu$m and 90~$\mu$m. The
terahertz radiation induces free carrier absorption in the lowest
conduction subband $e1$ because the photon energy is smaller than
the subband separation and much larger than the $\bm k$-linear
spin splitting. The samples were irradiated along the growth
direction.

In order to vary the angle between the polarization vector of the
linearly polarized light and the magnetic field we placed a metal
mesh polarizer behind a crystalline quartz $\lambda/4$-plate.
After passing through the $\lambda/4$-plate initially linearly
polarized laser light became circularly polarized. Rotation of the
metal grid enabled us to ob-\-
\begin{widetext}
%
%%%%%%%%%%TABLE%%%%%%%%%%%%%%%%%%%%%%%%%%%%%%%%%%%
\begin{center}
{\setlength{\tabcolsep}{1.5mm}
\begin{table}
\begin{tabular}{|c|l|c|c|c|c|}
\hline & Structure & Mobility& Electron density\\
      &            & cm$^2$/V$\cdot$s   & cm$^{-2}$\\
\hline A1 & (001)-InAs single QW of 15 nm width& $\approx 3\cdot 10^5$  & $8 \cdot 10^{11}$   \\
\hline A2 & (001)-GaAs double QW of 9.0 and 10.8 nm  width& $1.4\cdot 10^5$  & $1.12 \cdot 10^{11}$  \\
\hline A3 & (001)-GaAs heterojunction  & $3.53\cdot 10^6$  & $1.08 \cdot 10^{11}$  \\
\hline A4 & (001)-GaAs heterojunction  & $3.5\cdot 10^6$  &  $1.1 \cdot 10^{11}$  \\
\hline A5 & (001)-GaAs multiple QW (30 QWs of 8.2 nm width)  & $2.57\cdot 10^4$  & $9.3 \cdot 10^{11}$  \\
\hline
\end{tabular}
\caption{Parameters for non-illuminated samples at $T=4.2$~K. }
\label{tsample}
\end{table}
}
\end{center}
%%%%%%%%%%%%%%%%%%%%%%%%%%%%%%%%%%%%%%%%%%%%%%%%%%%%%%%%%%%%%%%%%%%%%%%
%
\end{widetext}
tain linearly polarized radiation with angle $\alpha=0^\circ \div
360^\circ$ between the $x'$ axis and the plane of linear
polarization of the light incident upon the sample.

To obtain elliptically and, in particular, circularly polarized
radiation the mesh polarizer behind the quartz $\lambda/4$-plate
was removed. The helicity $P_{\rm circ}$ of the incident light was
varied by rotating the  $\lambda/4$-plate according to $P_{\rm
circ} = \sin{2 \varphi}$ as given by Eq.~(\ref{circ}). For
$\varphi = n \cdot \pi/2$ with integer $n$ the radiation was
linearly polarized. Circular polarization was achieved with
$\varphi = (2n+1) \cdot (\pi /4)$, where even values of $n$
including $n=0$ yield the right-handed circular polarization
$\sigma_+$ and odd $n$ give the left-handed circular polarization
$\sigma_-$.

The photocurrent $\bm j$ was measured at room temperature in
unbiased structures via the voltage drop across a 50~$\Omega$ load
resistor in closed circuit configuration. The voltage was measured
with a storage oscilloscope. The measured current pulses of 100~ns
duration reflected the corresponding laser pulses.

\section{Experimental results} \label{experim}

As follows from Eqs.~(\ref{phen}), the most suitable experimental
arrangement for independent investigation of different
contributions to the magneto-induced photogalvanic effect is
achieved by applying magnetic field along one of the
crystallographic axes $x^\prime  \,\|\, [1{\bar 1}0]$, $y^\prime
\,\|\, [110]$ and measuring the in-plane current along or normal
to the magnetic field direction. Then, currents flowing
perpendicular to the magnetic field, contain contributions
proportional only to the parameters $S_1$ and $S_2$ if ${\bm B}
\parallel y'$ (or $S'_1$ and $S'_2$ if ${\bm B}
\parallel x'$), whereas, currents flowing parallel to the magnetic field
arise only from terms proportional to  $S_3$ and $S_4$ (or $S'_3$
and $S'_4$). Further separation of contributions may be obtained
by making use of the difference in their polarization
dependencies. The results obtained for $\lambda$ = 90~$\mu$m and
$\lambda$ = 148~$\mu$m  are qualitatively the same. Therefore
 we present only  data obtained for  $\lambda$ = 148~$\mu$m.

\subsection*{4.1. Photocurrent parallel to the magnetic field
($\bm j \,  \| \bm B \,\|\,y^\prime \,\| \,[110]$)}

According to Eqs.~(\ref{phen}) and Table~\ref{t5} only two
contributions  proportional to $S'_3$ and $S'_4$ are allowed in
this configuration. While the $S'_3$ contribution results in a
current for linear or elliptical polarization, the $S'_4$ one
vanishes for linear polarization and assumes its maximum at
circular polarization.

\begin{figure}
\centerline{\epsfxsize=0.85\linewidth \epsfbox{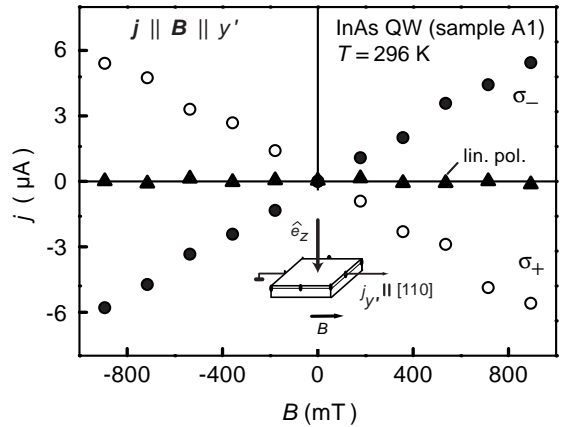}}
\caption{Magnetic field dependence of the photocurrent measured in
sample A1 at room temperature with the magnetic field $\bm B$
parallel to the $y'$ direction. Normally incident optical
excitation of $P \approx 4$~kW is performed at wavelength $\lambda
= 148\:\mu$m with {\em linear} ($\bm{E}\,\|\,x'$), right-handed
{\em circular} ($\sigma_+$), and left-handed {\em circular}
($\sigma_-$) polarization. The measured current component is {\em
parallel} to $\bm B$. The inset shows the experimental geometry.}
\label{fig1f}
\end{figure}

\begin{figure}
\centerline{\epsfxsize=0.85\linewidth \epsfbox{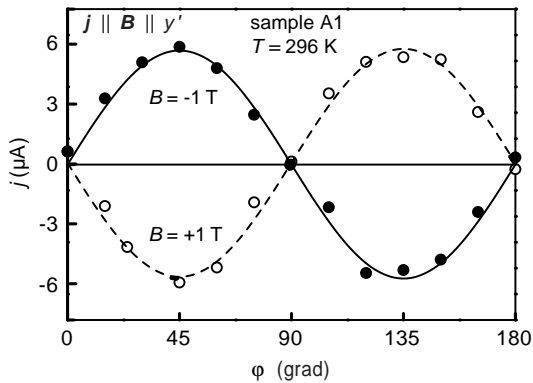}}
\caption{Photocurrent  as a function of the phase angle $\varphi$
defining the helicity. The photocurrent signal is measured in
sample A1 at room temperature in the configuration $\bm j \, \|
\bm B \,\|\,y^\prime$ for two opposite directions of the magnetic
field under normal incidence of the radiation with $\lambda =
148$~$\mu$m ($P \approx 4$~kW). The broken and full lines are
fitted after Eq.~\protect(\ref{circ}).} \label{fig2f}
\end{figure}

Irradiation of the samples A1$-$A4 subjected to an in-plane
magnetic field with normally incident {\em linearly} polarized
radiation cause no photocurrent. However, {\em elliptically}
polarized light yields  a helicity dependent
 current. Typical magnetic field and helicity dependencies
of this current are shown in Figs.~\ref{fig1f} and ~\ref{fig2f}.
The polarity of the current changes upon reversal of the applied
magnetic field as well as upon changing the helicity from right-
to left-handed. The polarization behavior of the current is well
described by $j_{y^{\prime}} \propto I B_{y^{\prime}} P_{circ}$.
This  means that the current is dominated by the last term on the
right side of the second equation~(\ref{phen}) (parameter $S'_4$)
while the third term is vanishingly small. Observation of a
photocurrent proportional to $P_{\rm circ}$ has already been
reported previously. This is the spin-galvanic
effect~\cite{Nature02}. The effect is caused by the optical
orientation of carriers, subsequent Larmor precession of the
oriented electronic spins and asymmetric spin relaxation
processes. Though, in general, the spin-galvanic current does not
require an application of magnetic field, it may  be considered as
a magneto-photogalvanic effect under the above experimental
conditions.

\begin{figure}
\centerline{\epsfxsize=0.85\linewidth \epsfbox{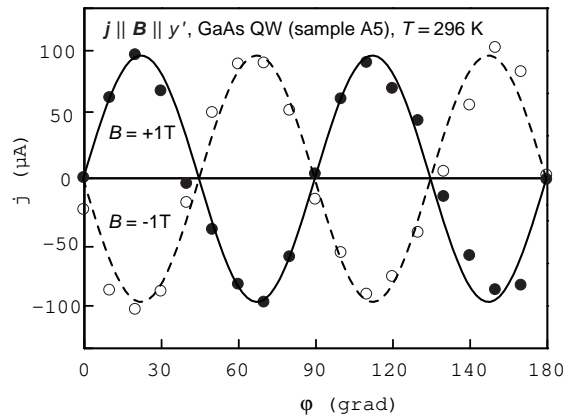}}
\caption{Photocurrent in the sample A5  as a function of the phase
angle $\varphi$ defining the helicity for magnetic fields of two
opposite directions. The photocurrent excited by normally incident
radiation of $\lambda = 148$~$\mu$m ($P \approx 17$~kW) is
measured at room temperature, $\bm j \, \| \bm B \,\|\,y^\prime$.
The broken and full lines are fitted after
Eq.~(\protect\ref{plin}).}
 \label{fig3f}
\end{figure}

\begin{figure}
\centerline{\epsfxsize=0.85\linewidth \epsfbox{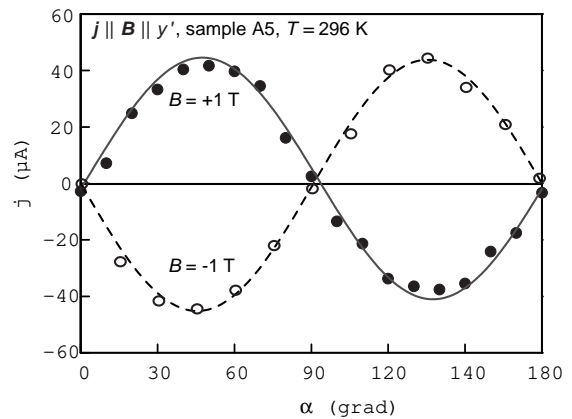}}
\caption{Photocurrent  in the sample A5  as a function of the
azimuth angle $\alpha$. The photocurrent $\bm j \, \| \bm B
\,\|\,y^\prime$ excited by normally incident {\em linearly}
polarized radiation of $\lambda = 148$~$\mu$m ($P \approx 17$~kW)
and measured at room temperature. The broken and full lines are
fitted according to Table~\protect\ref{t5}, 3$^{rd}$ term. }
\label{fig4f}
\end{figure}

One of our QW structures, sample A5, showed a quite different
behavior. In this sample the dependence of the magneto-induced
photocurrent on $\varphi$ is well described by $j_{y'} \propto
IB_{y'}{\rm sin}\,4\varphi$ (see Fig.~\ref{fig3f}). In contrast to
the samples A1$-$A4, in the sample A5 the spin-galvanic effect is
overweighed by the contribution of  the third term in
Eqs.~(\ref{phen}). The latter should also appear under excitation
with linearly polarized radiation. Figure~\ref{fig4f} shows the
dependence of the photocurrent on the angle $\alpha$ for one
direction of the magnetic field. The current $j_{y'}$ is
proportional to $IB_{y'}{\rm sin}\,2\alpha$ as expected for the
third term in Eqs.~(\ref{phen}).

\subsection*{4.2. Current perpendicular  to the magnetic field
($\bm j \,  \bot \bm B \,\|\,y^\prime \,\|\, [110]$)}

In the transverse geometry only contributions  proportional to the
parameters $S_1$ and $S_2$ are allowed. Here the samples A1 to A4
and  A5 again show different behavior.

\begin{figure}
\centerline{\epsfxsize=0.85\linewidth \epsfbox{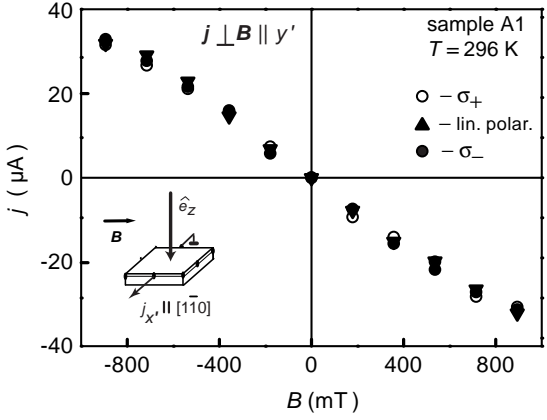}}
\caption{Magnetic field dependence of the photocurrent measured in
sample A1 at room temperature with the magnetic field $\bm B$
parallel to the $y'$ axis. Data are given for normally incident
optical excitation of $P \approx 4$~kW at the wavelength $\lambda
= 148\:\mu$m for {\em linear} ($\bm{E} \, \| \, x'$), right-handed
{\em circular} ($\sigma_+$), and left-handed {\em circular}
($\sigma_-$) polarization. The current is measured in the
direction {\em perpendicular} to $\bm B$.} \label{fig6f}
\end{figure}

The data of a magnetic field induced photocurrent perpendicular to
$\bm B$  in samples A1$-$A4  are illustrated in Fig.~\ref{fig6f}.
The  magnetic field dependence for sample A1 is shown for three
different polarization states. Neither rotation of the
polarization plane of the linearly polarized radiation nor
variation of helicity  changes the signal magnitude. Thus we
conclude that the current strength and sign are independent of
polarization. On the other hand, the current changes its direction
upon the magnetic field reversal. This behavior is described by
$j_{x^{\prime}} \propto I B_{y^{\prime}}$ and corresponds to the
first term on the right hand side of the first equation
in Eqs.~(\ref{phen}). The absence of a $\varphi$-dependence
indicates that the second term in Eqs.~(\ref{phen}) is negligibly
small. Note, that the dominant contribution to the  polarization
independent magneto-photogalvanic effect, described by the first
term on the right side of Eqs.~(\ref{phen}), is observed for the
same set of samples (A1$-$A4) where the longitudinal photocurrent
is caused by the spin-galvanic effect.

\begin{figure}
\centerline{\epsfxsize=0.85\linewidth \epsfbox{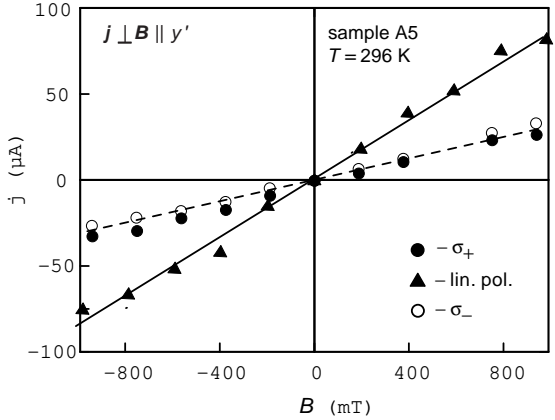}}
\caption{Magnetic field dependence of the photocurrent measured in
sample A5 at room temperature with the magnetic field $\bm B$
parallel to the $y'$ axis. Data are presented for normally
incident optical excitation $P \approx 17$~kW at the wavelength
$\lambda = 148\:\mu$m for the {\em linear} ($\bm{E} \, \| \, x'$),
right-handed {\em circular} ($\sigma_+$), and left-handed {\em
circular} ($\sigma_-$) polarization. The current is measured in
the direction {\em perpendicular} to~$\bm B$. } \label{fig7f}
\end{figure}

\begin{figure}
\centerline{\epsfxsize=0.85\linewidth \epsfbox{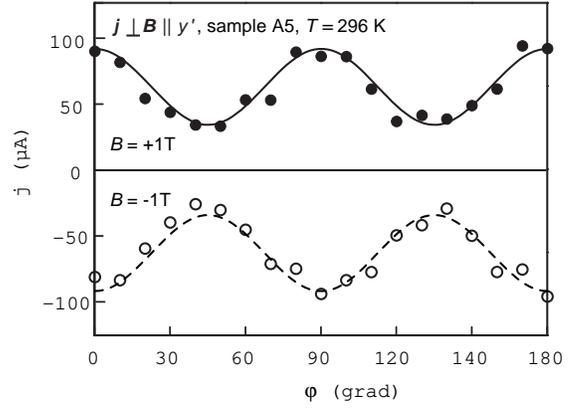}}
\caption{Photocurrent in  sample A5 as a function of the phase
angle $\varphi$ defining the Stokes parameters, see Eq.
(\ref{plinprime}). The photocurrent excited by normally incident
radiation of $\lambda = 148$~$\mu$m 9$P \approx 17$~kW)is measured
at room temperature, $\bm j \, \bot \bm B \,\|\,y^\prime$. The
full and broken lines are fitted according to Table \protect
\ref{t5}, the 1$^{st}$ and 2$^{nd}$ terms.
 } \label{fig8f}
\end{figure}

\begin{figure}
\centerline{\epsfxsize=0.85\linewidth \epsfbox{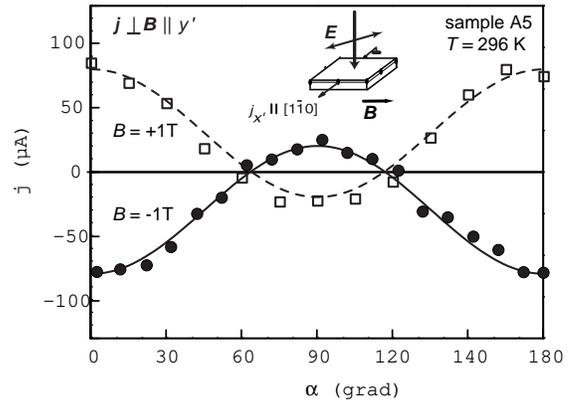}}
\caption{Photocurrent  in sample A5 for $\bm j \, \bot \bm B
\,\|\,y^\prime$ as a function of the azimuth angle $\alpha$. The
photocurrent excited by normally incident radiation of $\lambda =
148$~$\mu$m ($P \approx 17$~kW) is measured at room temperature
for magnetic fields of two opposite directions. The broken and
full lines are fitted according to Table~ \protect \ref{t5}, the
1$^{st}$ and 2$^{nd}$ terms.} \label{fig9f}
\end{figure}

In sample A5 a clear polarization dependence, characteristic for
the second terms in Eqs.~(\ref{phen}), has been detected. The
magnetic field and the polarization dependencies obtained from
this sample are shown in Figs.~\ref{fig7f}, \ref{fig8f} and
\ref{fig9f}, respectively. For the sample A5 the
$\varphi$-dependence can be well fitted  by $S_1+ S_2(1+{\rm
cos}\,4\varphi)/2$ while the $\alpha$-dependence is $S_1 +
S_2\,{\rm cos}\,2\alpha$, as expected for the first and second
terms in Eqs.~(\ref{phen}).

\subsection*{4.3. Magnetic field applied along the $x^\prime \parallel [1\bar1 0]$ direction}

Rotation of $\bm B$ by 90$^\circ$ with respect to the previous
geometry interchanges  the role of the axes $x^\prime$ and
$y^\prime$. Now the magnetic field is applied along the
$[1\bar{1}0]$ crystallographic direction. The magnetic field and
polarization dependencies observed experimentally in both
configurations are qualitatively similar. The only difference is
the magnitude of the photocurrent.
%For the samples A1$-$A4 the
%photocurrent components both parallel and perpendicular to the
%magnetic field are smaller than that for $\bm{B}\parallel
%y^\prime$.
The observed difference in photocurrents is expected
for $C_{2v}$ point symmetry of the QW where the axes $[1 \bar{1}
0]$ and $[110]$ are non-equivalent. This is taken into account in
Eqs.~(\ref{phen}) by introducing independent parameters $S_i$ and
$S^\prime_i$ ($i = 1 \dots 4$).

\subsection*{4.4. Magnetic field applied along the crystallographic
axis $x \parallel$ [100]}

\begin{figure}[ht]
\centerline{\epsfxsize=0.85\linewidth \epsfbox{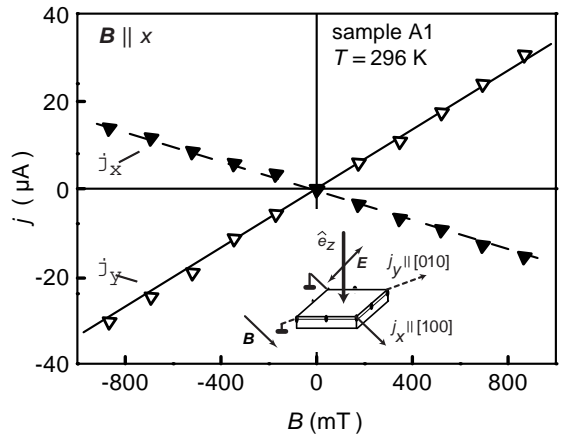}}
\caption{Magnetic field dependence of the photocurrent measured in
sample A1 with the magnetic field $\bm B$ parallel to the [100]
axis under photoexcitation with normally incident light of the
wavelength $\lambda = 148\:\mu$m ($P \approx 4$~kW) for {\em
linear} polarization $\bm{E} \, \| \, y$. The current is measured
in the directions {\em parallel} ($j_x$) and {\em perpendicular}
($j_y$) to $\bm B$.} \label{fig10f}
\end{figure}

\begin{figure}[ht]
\centerline{\epsfxsize=0.85\linewidth \epsfbox{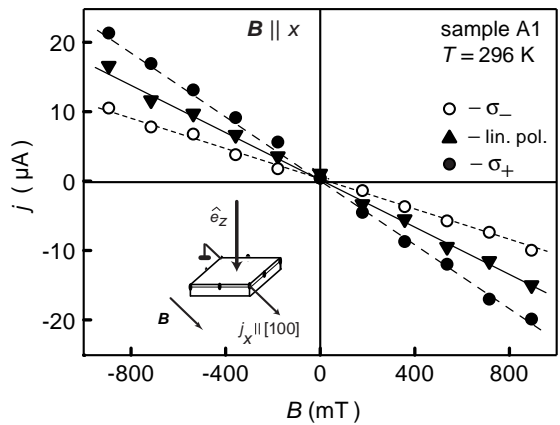}}
\caption{Magnetic field dependence of the photocurrent measured in
sample A1 with the magnetic field $\bm B$ parallel to the [100]
axis. Optical excitation of $P \approx 4$~kW at normal incidence
was applied at wavelength $\lambda = 148\:\mu$m for {\em linear}
($\bm{E} \, \| \, y$), right-handed {\em circular} ($\sigma_+$),
and left-handed {\em circular} ($\sigma_-$) polarization. The
current is measured in the direction {\em parallel} to $\bm B$.}
\label{fig11f}
\end{figure}

Under application of $\bm B$ along one of the in-plane cubic axes
in a (001)-grown structure, all contributions to the photocurrent
are allowed. This can be seen from Eqs.~(\ref{phena}) and
Table~\ref{t2}. In all samples both longitudinal and transverse
currents are observed for  {\em linearly} (Fig.~\ref{fig10f}) as
well as {\em circularly} (Fig.~\ref{fig11f}) polarized excitation.
In the absence of the magnetic field the current signals vanish
for all directions. For the samples A1$-$A4 a clear spin-galvanic
current proportional to helicity $P_{circ}$ and superimposed on a
helicity independent contribution is detected (see
Fig.~\ref{fig11f}). The possibility of extracting the
spin-galvanic effect is of particular importance in experiments
aimed at the  separation of Rashba- and Dresselhaus-like
contributions to the spin-orbit interaction as has been recently
reported~\cite{PRL04BIASIA}.


%%%%%%%%%%%%%%%%%%%%%%%%%%%%%%%%%%%%%%%%%%%%%%%%%%%%%%%%%%%%%%%%

\section{Microscopic models}
\label{micmodel}

The term magneto-photogalvanic effects (MPGE) stands for  the
generation of  magnetic field induced photocurrent under polarized
or unpolarized optical excitation. In this Section we give a
survey of possible microscopic mechanisms leading to  MPGE.
Besides mechanisms discussed in literature we also present here
novel mechanisms. We start by recalling non-gyrotropic
spin-independent mechanisms used to interpret MPGE observed in
bulk non-centrosymmetric semiconductors (Section~5.1). They are
based on the cyclotron motion of free carriers in both the real
and  the ${\bm k}$-space. Since in a QW subjected to an in-plane
magnetic field, the cyclotron motion is suppressed  one needs to
seek for alternative mechanisms. As we will  demonstrate below
(Sections 5.3 to 5.5), the generation of magneto-induced
photocurrent in quantum wells  requires both  gyrotropy and
magnetic field  and therefore the effects belong to the
magneto-gyrotropic class.

\subsection*{ 5.1. Bulk semiconductors of the T$_d$ point
symmetry} \label{micmodel1}

In this Section we outline briefly microscopic mechanisms
responsible for magneto-photocurrents generated in bulk materials
of the T$_d$ symmetry.

\paragraph*{Non-gyrotropic, spin-independent mechanisms.}

The phenomenological description of the MPGE in the T$_d$-class
bulk crystals are described by Eqs.~(\ref{phend})$-$(\ref{phen3})
in Appendix A. Microscopically, the contribution proportional to
$S_2$ in Eq.~(\ref{phend}) can be easily
interpreted~\cite{ferro,ivlyapi} as the Hall rotation of the
zero-magnetic field photocurrent. At zero magnetic field  the
current ${\bm j}^{(0)}$ in response to linear polarized radiation
is given by
$$
j_x^{(0)} \propto e_y e_z^* + e_z e_y^*\:,\;\;j_y^{(0)} \propto
e_z e_x^* + e_x e_z^*\:,\;\; j_z^{(0)} \propto e_x e_y^* + e_x
e_y^*\:.
$$
Applying a magnetic field ${\bm B}$ yields a current  ${\bm j}$ in
the direction parallel to the vector ${\bm B} \times {\bm
j}^{(0)}$. The coefficient $S_1$, on the other hand, determines
the contribution to the photocurrent arising even if ${\bm
j}^{(0)} = 0$, e.g., for ${\bm e}
\parallel x$. This particular contribution can be described microscopically
as follows \cite{ivpira} (see also \cite{lyanda1,andryanov1}): (a)
optical alignment of free-carrier momenta described by an
anisotropic correction to the free-carrier non-equilibrium
distribution function, $\delta f({\bm k})$, proportional to
$k_{\alpha} k_{\beta}/k^2$; (b) new terms $k_{\gamma}
k_{\delta}/k^2$ appear due to cyclotron rotation of the
free-carrier distribution function; (c) momentum scattering of
free carriers results in an electric current $j_{\eta} \propto
C_{\eta + 1,\, \eta + 2}$, where $\eta = (1,2,3) \equiv (x,y,z)$,
$C_{\gamma, \,\delta}$ are the coefficients in the expansion of
$\delta f({\bm k})$ over $k_{\gamma} k_{\delta}/k^2$. Here,  the
cyclic permutation of indices is assumed. The current appears
under one-phonon induced free carrier shifts in the real space
(the so-called shift contribution) or due to two-phonon asymmetric
scattering (the ballistic contribution)~\cite{book,sturman}. For
the polarization ${\bm e}
\parallel x$, the anisotropic part of the free-carrier
non-equilibrium distribution function is proportional to
$k_x^2/k^2$. For ${\bm B} \parallel y$, the cyclotron rotation of
this anisotropic distribution leads to the term $\delta f({\bm k})
\propto k_x k_z/k^2$. The further momentum relaxation yields an
electric current in the $y$ direction. It should be mentioned that
a similar mechanism contributes to $S_2$. It is clear that both
this mechanism and the photo-Hall mechanism are spin-independent
since the free-carrier spin is not involved here. Note that both
mechanisms do exist in  bulk crystals of the T$_d$ symmetry which
are non-gyrotropic. Therefore they can be classified  as {\it
non-gyrotropic} and {\it spin-independent}.

An important point to stress is that the above mechanisms vanish
in QWs for an in-plane magnetic field. Because the free-carrier
motion  is quantized in growth direction the anisotropic
correction $\delta f({\bm k}) \propto k_{\eta} k_z/k^2$ ($\eta =
x, y$) to the distribution function does not exist.

\paragraph*{Non-gyrotropic, spin-dependent mechanisms.}

Two non-gyrotropic but spin-dependent me\-cha\-nisms causing
magnetic field induced photocurrents were proposed for bulk
zinc-blende-lattice semiconductors in \cite{magarill1,emelya}. In
\cite{magarill1} the photocurrent is calculated for optical
transitions between spin-split Landau-level subbands under
electron spin resonance conditions in the limit of strong magnetic
field. Taking into account both the spin-dependent Dresselhaus
term, cubic in the wavevector ${\bm k}$,
\begin{equation} \label{k3}
{\cal H}^{(3)}({\bm k}) = \gamma [ \sigma_x k_x (k_y^2 - k_z^2) +
\sigma_y k_y (k_z^2 - k_x^2) + \sigma_z k_z (k_x^2 - k_y^2)]
\end{equation}
and the quadratic in ${\bm k}$ Zeeman term
\begin{equation} \label{bk2}
{\cal H}^{(2)}({\bm B}) = {\cal G} ({\bm \sigma} \cdot {\bm
k})({\bm B} \cdot {\bm k})
\end{equation}
in the bulk electron Hamiltonian, spin-flip optical transitions
lead to asymmetric photoexcitation of electrons in the ${\bm
k}$-space and, hence, to a photocurrent. At a fixed radiation
frequency the photocurrent has a resonant nonlinear dependence on
the magnetic field and contains contributions both even and odd as
a function of $\bm B$. In Ref.~\cite{emelya} the photocurrent
under impurity-to-band optical transitions in bulk InSb was
described taking into account the quantum-interference of
different transition channels one of which includes an
intermediate intra-impurity spin-flip process. This photocurrent
is proportional to  photon momentum and depends on the light
propagation direction. Therefore, it can be classified as the
photon drag effect which occurs under impurity-to-band optical
transitions and is substantially modified by the intra-impurity
electron spin resonance. Since in the present work the experiments
were performed under normal incidence of radiation of two
dimensional structure we will not consider the photon drag effect
in the following discussion.

\subsection*{ 5.2. Effects of gyrotropy in (001)-grown quantum wells}
\label{gyrotropy}

The (001)-grown quantum well structures are characterized by a
reduced symmetry D$_{2d}$ (symmetric QWs) or C$_{2v}$ (asymmetric
QWs). Generally, for symmetry operations of these point groups,
the in-plane components of a polar vector ${\bm R}$ and an axial
vector ${\bm L}$ transform according to the same representations.
In the C$_{2v}$ group there are two invariants which can be
constructed from the products $R_{\alpha} L_{\beta}$, namely,
\begin{eqnarray} \label{i1i2}
{\cal I}_1 &=& R_x L_x - R_y L_y = R_{x'} L_{y'} + R_{y'}
L_{x'}\:,
\\ {\cal I}_2 &=& R_x L_y - R_y L_x = R_{x'} L_{y'} - R_{y'} L_{x'}
\equiv ({\bm R} \times {\bm L})_z \:. \nonumber
\end{eqnarray}
The D$_{2d}$ symmetry allows only one invariant, ${\cal I}_1$. In
the following ${\cal I}_1$- and ${\cal I}_2$-like functions or
operators are referred to as the gyrotropic invariants.

In order to verify that a given function, ${\cal I}(\bm{k}',
\bm{k})$, linear in $\bm{B}$ or $\bm{\sigma}$ contains a
gyrotropic invariant one can use a simple criterion, namely,
multiply ${\cal I}$ by $k_{\eta}$ and $k'_{\eta}$ ($\eta = x, y$),
average the product over the directions of $\bm{k}'$ and $\bm{k}$
and check that the average is nonzero. Three examples of
gyrotropic invariants relevant to the present work are given
below.

The first is the  spin-orbit part of the electron effective
Hamiltonian,
\begin{eqnarray}\label{siabia}
{\cal H}_{\rm BIA}^{(1)} &=& \beta_{\rm BIA} (\sigma_x k_x -
\sigma_y k_y) \:, \\
{\cal H}_{\rm SIA}^{(1)} &=& \beta_{\rm SIA} (\sigma_x k_y -
\sigma_y k_x)\:, \nonumber \\
{\cal H}_{\rm BIA}^{(3)} &=& \gamma_{\rm BIA} (\sigma_x k_x k^2_y
- \sigma_y k_y k^2_x)\:, \nonumber \\
{\cal H}_{\rm SIA}^{(3)} &=& \gamma_{\rm SIA} (\sigma_x k_y -
\sigma_y k_x)k^2 \:. \nonumber
\end{eqnarray}
Here $\sigma_{\alpha}$ are the spin Pauli matrices, $k_x$ and
$k_y$ are the components of the 2D electron wavevector,
$\gamma_{\rm BIA}$ coincides with the parameter $\gamma$
introduced by Eq.~(\ref{k3}), ${\cal H}^{(1)}_{\rm BIA}$ and
${\cal H}^{(1)}_{\rm SIA}$ are the so-called Dresselhaus and
Rashba terms being linear in $\bm{k}$  or, respectively, bulk inversion
asymmetry (BIA) and structure inversion asymmetry (SIA) terms.
The  terms ${\cal
H}^{(1)}_{\rm BIA}$ and ${\cal H}_{\rm BIA}^{(3)}$, linear and cubic in $\bm{k}$ , stem from
averaging the cubic-$\bm{k}$ spin-dependent Hamiltonian
Eq.~(\ref{k3}).

The second example of a gyrotropic invariant is the well known
diamagnetic band shift existing in  asymmetric
QWs~\cite{stern,ando1,andostern}, see also
\cite{kotthaus,zawadzki,zaremba}. This spin-independent
contribution to the electron effective Hamiltonian reads
\begin{equation} \label{diamaH}
{\cal H}^{ \rm dia}_{\rm SIA} = \tilde{\alpha}_{\rm SIA} (B_x k_y - B_y
k_x)\:.
\end{equation}
The coefficient $\tilde{\alpha}_{\rm SIA}$ in the $\nu$-th
electron subband is given by $\tilde{\alpha}_{\rm SIA}^{(\nu)} = (
e \hbar /c m^*) \bar{z}_{\nu}$, where $m^*$ is the effective
electron mass, and $\bar{z}_{\nu} = \langle e \nu \vert z \vert e
\nu \rangle$ is the center of mass of the electron envelope
function in this subband.

The last example is an asymmetric part of electron-phonon
interaction. In contrast to the previous two examples it does not
modify the single-electron spectrum but can give rise to spin
dependent effects. It leads, e.g., to spin photocurrents
considered in Sections 5.3 and 5.4.
The electron-phonon interaction is given by
\begin{equation}\label{scattering}
\hat{V}_{\rm el-phon}(\bm{k'}, \bm{k})= \Xi_c \, \sum_j
\epsilon_{jj} + \Xi_{cv} \xi \sum_j [(\bm{k}' + \bm{k}) \times
\bm{\sigma}]_j\ \epsilon_{j+1\,j+2}\:.
\end{equation}
Here $\epsilon_{j j'}$ is the phonon-induced strain tensor
dependent on the phonon wavevector $\bm{q} = \bm{k}' - \bm{k}$,
$\Xi_c$ and $\Xi_{cv}$ are the intra- and inter-band constants of
the deformation potential. For zinc-blende-lattice QWs the
coefficient $\xi$ is given by~\cite{IT_jetp}
\begin{equation}
\xi = \frac{i \hbar p_{cv}}{3 m_0}
\frac{\Delta_{so}}{\varepsilon_g (\varepsilon_g+\Delta_{\rm
so})}\:,
\end{equation}
where $m_0$ is the free-electron mass, $\varepsilon_g$ and
$\Delta_{\rm so}$ are the band gap and the valence band spin-orbit
splitting of the bulk semiconductor used in the QW layer, $p_{cv}
= \langle S | \hat{p}_z | Z \rangle$ is the interband matrix
element of the momentum operator between the Bloch functions of
the conduction and valence bands, $S$ and $Z$.

Compared with the non-gyrotropic class T$_d$ the presence of
gyrotropic invariants in the electron effective Hamiltonian in QWs
of the D$_{2d}\,$- and C$_{2v}\,$-symmetry enable new mechanisms
of the MPGE. At present we are unaware of any non-gyrotropic
mechanism of the MPGE in QW structures in the presence of an
in-plane magnetic field. Thus, it is natural to classify such
contributions to the MPGE as {\it magneto-gyrotropic
photocurrents}. Below we consider microscopic mechanisms of
magneto-gyrotropic photocurrents, both {\it spin-dependent} and
{\it spin-independent}. To illustrate them we present model
pictures for three different mechanisms connected to acoustic
phonon assisted optical transitions. Optical phonon- or
defect-assisted transitions and those involving electron-electron
collisions may be considered in the same way.

\subsection*{ 5.3. Photocurrent due to spin-dependent asymmetry of
optical excitation} \label{asexcitation}

The first possible mechanism of current generation in QWs in the
presence of a magnetic field is related to the asymmetry of
optical excitation. The characteristic feature of this mechanism
is a sensitivity to the  polarization of light. In our experiments
we employ free-electron absorption. Indirect optical transitions
require a momentum transfer from phonons to electrons. A
photocurrent induced by these transitions appears due to an
asymmetry of either electron-pho$t$on or electron-pho$n$on
interaction in the $\bm k$-space. Below we take into account the
gyrotropic invariants within the first order of the perturbation
theory. Therefore while considering the spin-dependent
magneto-gyrotropic effects, we can replace the contribution to the
electron Hamiltonian linear in the Pauli spin matrices by only one
of the terms proportional to the matrix $\sigma_j$ and perform the
separate calculations for each index $j$. Then spin-conserving and
spin-flip mechanisms can be treated independently.

\begin{figure}
\centerline{\epsfxsize=0.75\linewidth \epsfbox{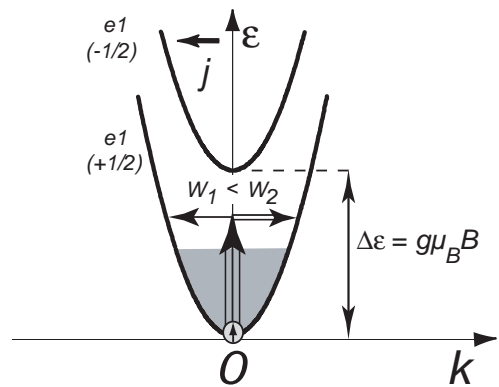}}
\caption{Microscopic origin of photocurrent caused by asymmetric
photoexcitation in an in-plane magnetic field. The spin subband
$(+1/2)$ is preferably occupied due to the Zeeman splitting. The
rates of optical transitions for opposite wavevectors $ k$ are
different, $W_1<W_2$. The $\bm{k}$-linear spin splitting is
neglected in the band structure because it is unimportant for this
mechanism.} \label{model1}
\end{figure}

\paragraph*{5.3.1. Spin-dependent spin-conserving asymmetry of
photoexcitation due to asymmetric electron-phonon interaction.}
\label{1.1.1.}

In gyrotropic media the electron-phonon interaction $\hat{V}_{\rm
el-phon}$ contains, in addition to the main contribution, an
asymmetric spin-dependent term $\propto \sigma_{\alpha} (k_{\beta}
+ k_{\beta}')$  given by Eq.~(\ref{scattering}), see
also~\cite{bulli,Belinicher,AGW,IT_jetp}. Microscopically this
contribution is caused by  structural and bulk inversion asymmetry
alike  Rashba/Dresselhaus band spin splitting in the
$\bm{k}$-space. The asymmetry of  electron-phonon interaction
results in non-equal rates of indirect optical transitions for
opposite wavevectors in each spin subband with $s_{\alpha} = \pm
1/2$. This causes an asymmetric distribution of photoexcited
carriers within the subband $s_{\alpha}$ and yields therefore  a
flow, ${\bm i}_{\alpha}$, of electrons in this subband. This
situation is sketched in Fig.~\ref{model1} for the spin-up
$(s=1/2)$ subband. The single and double horizontal arrows in
Fig.~\ref{model1} indicate the difference in electron-phonon
interaction strength for positive and negative wavevectors. The
important point now is that single and double arrows are
interchanged for the other spin direction (see
Eq.~(\ref{scattering})). Indeed the enhancement of the
electron-phonon interaction rate for a specific $k$-vectors
depends on the spin direction. Therefore for the other spin
subband, the situation is reversed. This is analogous to the well
known spin-orbit interaction where the shift of the
$\varepsilon({\bm k})$ dispersion depends also on the spin
direction. Thus without magnetic field two oppositely directed and
equal currents in spin-up and spin-down subbands cancel each other
exactly. This non-equilibrium electron distribution in the ${\bm
k}$-space is characterized by zero electric current but nonzero
pure spin current ${\bm i}_{\rm spin}$ = $(1/2) ({\bm i}_{1/2} -
{\bm i}_{- 1/2})$~\cite{IT}. The application of a magnetic field
results, due to the Zeeman effect, in    different equilibrium
populations of the subbands. This is seen in Fig.~\ref{model1},
where the Zeeman splitting is largely exaggerated to simplify
visualization. Currents flowing in opposite directions become
non-equivalent resulting in a spin polarized net electric current.
Since the current is caused by asymmetry of photoexcitation, it
may depend on the polarization of radiation.

Generally, indirect optical transitions are treated in
perturbation theory as second-order processes involving virtual
intermediate states. The compound matrix element of
phonon-mediated transition $(s, \bm{k}) \to (s', \bm{k}')$ with
the intermediate state in the same subband $e1$ can be written as
\begin{equation} \label{indirm}
M^{(\pm)}_{s' \bm{k}', s \bm{k}} =
\end{equation}
\vspace{-0.8cm}
\[
\sum_{s''} \left[ \frac{V^{(\pm)}_{s' \bm{k}', s'' \bm{k}} R_{s'',
s}({\bm k})}{\varepsilon_{s}(\bm{k}) - \varepsilon_{s''}(\bm{k}) +
\hbar \omega} + \frac{ R_{s', s''}({\bm k}') V^{(\pm)}_{s''
\bm{k}', s \bm{k}}}{\varepsilon_{s}(\bm{k}) -
\varepsilon_{s''}(\bm{k}') \mp \hbar \Omega({\bm q})} \right] \:,
\]
where $R_{s', s}({\bm k})$ is the direct optical matrix element,
$V^{(\pm)}_{s' \bm{k}', s \bm{k}}$ is the matrix element of
phonon-induced scattering, the upper (lower) sign in $\pm$ and
$\mp$ means the indirect transition involving absorption
(emission) of a phonon; $s, s'$ and $s''$ are the spin indices.

While considering the spin-conserving electron transitions, we use
the basis of electron states with the spin components $s = \pm
1/2$ parallel to the direction $\eta
\parallel {\bm B}$, retain in the gyrotropic invariants only the
spin-independent terms  containing $\sigma_{\eta}$  and consider
the processes $(s, \bm{k}) \to (s, \bm{k}')$. Then, in
Eq.~(\ref{indirm}) one can set $s = s' = s''$ and reduce the
equation to
\begin{equation}
M^{(\pm)}_{s \bm{k}', s \bm{k}} = V^{(\pm)}_{s \bm{k}', s \bm{k}}
\ [ R_{s, s}({\bm k}) -  R_{s, s}({\bm k}') ] / \hbar \omega \:.
\end{equation}
The photocurrent density is given by
\begin{widetext}
\begin{equation} \label{indirj}
\bm{j} = e \frac{2 \pi}{\hbar} \sum_{\bm{k}' \bm{k} s \pm}
[\bm{v}_{s}(\bm{k}') \tau'_p - \bm{v}_{s}(\bm{k}) \tau_p ]\
|M^{(\pm)}_{s \bm{k}', s \bm{k}}|^2 \{f^0_s(\bm{k}) [1 -
f^0_s(\bm{k}')] N^{(\pm)}_{\bm{q}} - f^0_s(\bm{k}') [1 -
f^0_s(\bm{k})] N^{(\mp)}_{\bm{q}}\}\ \delta [
\varepsilon_s(\bm{k}') - \varepsilon_s(\bm{k}) - \hbar \omega \pm
\hbar \Omega(\bm{q}) ] \:,
\end{equation}
\end{widetext}
where $e$ is the electron charge, $\bm{v}_s(\bm{k})$ is the
electron group velocity in the state $(s, \bm{k})$, $\tau_p$ and
$\tau'_p$ are the electron momentum relaxation times in the
initial and final states, $f^0_s(\bm{k})$ is the electron
equilibrium distribution function, $\bm{q} = \bm{k}'- \bm{k}$ is
the phonon wavevector, $N^{(\pm)}_{\bm{q}} = N_{\bm{q}} + (1 \pm
1)/2$, and $N_{\bm{q}}$ is the phonon occupation number.

For the mechanism in question one retains in $R_{s, s}({\bm k})$
the main contribution $- (e A_0/cm^*) (\hbar {\bm k} \cdot {\bm
e})$ and uses the electron-phonon interaction in the form of
Eq.~(\ref{scattering}) which can be rewritten as
\begin{equation}   \label{vscat}
V_{s \bm{k}', s \bm{k}} = \Xi_c \, \epsilon_{ii} + \Xi_{cv}\ \xi
[(\bm{k}' + \bm{k}) \times \bm{\sigma}_{ss}]_z\ \epsilon_{xy} \:.
\end{equation}
Here $A_0, {\bm e}$ are the scalar amplitude and polarization unit
vector of the light vector-potential, and $\epsilon_{ii} \equiv
\sum_i \epsilon_{ii}$.

Under indirect photoexcitation, the asymmetry of scattering
described by Eq.~(\ref{vscat}) leads to electric currents of
opposite directions in both spin subbands. The net electric
current occurs due to the Zeeman splitting induced selective
occupation of these branches in equilibrium. We remind that, in
the first order in the magnetic field ${\bm B}$, the average
equilibrium electron spin is given by
\begin{equation} \label{spineq}
{\bm S}^{(0)} = - \frac{g \mu_B {\bm B}}{4 \bar{\varepsilon}}\:,
\end{equation}
where $g$ is the electron effective $g$-factor, $\mu_B$ is the
Bohr magneton, $\bar{\varepsilon}$ is the characteristic electron
energy defined for the 2D gas as $\int d\varepsilon f(\varepsilon)
/ f(0)$, with $f(\varepsilon)$ being the equilibrium distribution
function at zero field, so that $\bar{\varepsilon}$ equals the
Fermi energy, $\varepsilon_F$, and the thermal energy, $k_B T$,
for degenerate and non-degenerate electron gas, respectively. The
current, induced by electron-phonon asymmetry under indirect
photoexcitation, can be estimated as
\[
j \propto \frac{e \tau_p}{\hbar} \frac{\Xi_{cv} \xi}{\Xi_{c}}
\,\eta_{ph} I S^{(0)} \:,
\]
where $\eta_{ph}$ is the phonon-assisted absorbance of the
terahertz radiation.

For impurity-assisted photoexcitation, instead of
Eq.~(\ref{vscat}), one can use the spin-dependent matrix element
of scattering by an impurity,
\begin{equation}   \label{vscatim}
V_{s \bm{k}', s \bm{k}} = \{ V_0({\bm q}) + V_z({\bm q})\ \xi
[(\bm{k}' + \bm{k}) \times \bm{\sigma}_{ss}]_z \} {\rm e}^{{\rm i}
(\bm{k} - \bm{k}') \bm{r}_{\rm im}}\:,
\end{equation}
where ${\bm q} = \bm{k}' - \bm{k}$, $V_0$ is the matrix element
for intra-band electron scattering by the defect, $V_z$ is the
matrix element of the defect potential taken between the
conduction-band Bloch function $S$ and the valence-band function
$Z$ (see~\cite{IT_jetp} for details ), $\bm{r}_{\rm im}$ is the
in-plane position of the impurity.

\paragraph*{5.3.2. Asymmetry of photoexcitation due to
asymmetrical electron-phonon spin-flip scattering.} \label{1.1.2.}
%
%%%*****
Indirect optical transitions involving  phonon-induced asymmetric
spin-flip scattering also lead to an electric current if spin
subbands get selectively occupied due to Zeeman splitting.
The
asymmetry can be due to a dependence of the spin-flip scattering
rate on the transferred wavevector $\bm{k}'-\bm{k}$ in the system
with the odd-$\bm{k}$ spin splitting of the electron subbands,
see~\cite{Nature02}. Estimations show that this mechanism  to
the photocurrent is negligible compared to the previous
mechanism~5.3.1.

\paragraph*{5.3.3. Spin-dependent spin-conserving asymmetry of
photoexcitation due to asymmetric electron-pho$t$on interaction.}
\label{1.1.3.}

A magnetic field induced photocurrent under linearly polarized
excitation can occur due to an asymmetry of electron-pho$t$on
interaction. The asymmetry is described by the optical matrix
element
\begin{equation} \label{spincm}
R_{s, s}({\bm k}) = -  \frac{e A_0}{c} \left[ \frac{\hbar ({\bm k}
\cdot {\bm e})}{m^*} + \frac{1}{\hbar}\sum_{j} e_j
\frac{\partial}{\partial k_j} {\cal H}^{(3)}_{ss}({\bm k}; \eta)
\right] \:,
\end{equation}
where $ {\cal H}^{(3)}_{ss}({\bm k}; \eta)$ is  the
$\sigma_{\eta}$-dependent term in the cubic-${\bm k}$ contribution
${\cal H}_{\rm BIA}^{(3)}({\bm k}) + {\cal H}_{\rm SIA}^{(3)}({\bm
k})$ to the electron Hamiltonian. Here, for the electron-phonon
matrix element, one can take the main spin-independent
contribution including both the piezoelectric and
deformation-potential mechanisms. Under  indirect light
absorption, the electron-photon asymmetry results in electric
currents flowing in opposite directions in both spin branches.
Similarly to the mechanism~5.3.1, the net electric current is
nonzero due to the selective occupation of the Zeeman-split spin
branches.

It should be stressed that the ${\cal H}^{(3)}_{ss}({\bm k};
\eta)$ term should also be taken into account in the
$\delta$-function, the distribution function and the group
velocity in the microscopical expression (\ref{indirj}) for the
photocurrent. Note that the linear-${\bm k}$ terms in the
effective electron Hamiltonian, see~Eq.~(\ref{siabia}), do not
lead to a photocurrent in the first order in $\beta_{\rm BIA}$ or
$\beta_{\rm SIA}$ because the linear-$k_i$ term in the function
$\hbar^2 k_i^2/2 m^* + \beta k_i$ disappears after the replacement
$k_i \to \tilde{k}_i = k_i + \beta m^*/\hbar^2$.

\paragraph*{5.3.4. Asymmetry of spin-flip photoexcitation due to
asymmetric electron-pho$t$on interaction.} \label{1.1.4.}

To obtain the asymmetric photoexcitation for optical spin-flip
processes we can take  into account, alongside with the term
odd-${\bm k}$, the quadratic-${\bm k}$ Zeeman term similar to that
introduced by Eq. (\ref{bk2}). Then the spin-flip optical matrix
element is given by
\begin{equation}\label{spinflipm}
R_{\bar{s}, s}({\bm k}) = -  \frac{e A_0}{\hbar c}
\end{equation}
\vspace{-1cm}
\[
\times \left\{ {\cal G} {\bm \sigma}_{\bar{s}, s} \cdot [ {\bm e}~
({\bm B} \cdot {\bm k}) +  {\bm k}~ ({\bm B} \cdot {\bm e})] +
\sum_{j} e_j \frac{\partial}{\partial k_j} {\cal H}_{\bar{s},
s}({\bm k}) \right\}\:,
\]
where $\bar{s} = - s$ and ${\cal H}({\bm k})$ is the odd-${\bm k}$
contribution to the electron Hamiltonian, including both linear
and cubic terms. Estimations show that the photocurrent due to the
spin-conserving processes described by Eq.~(\ref{spincm}) is
larger than that due to the spin-flip processes described by
Eq.~(\ref{spinflipm}).

\paragraph*{5.3.5. Spin-dependent asymmetry of indirect transitions via
other bands or subbands.} \label{1.1.5.}

This contribution is described by Eq.~(\ref{indirm}) where the
summation is performed over virtual states in subbands different
from $e1$. The estimation shows that it is of the same order of
magnitude as the contribution due to the mechanism~5.3.1.

Summarizing  the above mechanisms we would like to stress that the
characteristic feature of all of them is a sensitivity to the
light linear polarization described in Eqs.~(\ref{phen}) by the
terms proportional to $S_2, S'_2, S_3, S'_3$. Depending on the
particular set of parameters, e.g., those in Eqs.~(\ref{siabia},
\ref{scattering}), the energy dependence of $\tau_p$, the ratio
between the photon energy, the electron average energy etc., one
can obtain any value for the ratio between $S_2$ and $S_3$ as well
as for the ratio between one of them and the coefficient $S_1$.

\subsection*{ 5.4. Current due to spin-dependent asymmetry of electron
relaxation} \label{asrelaxation}

Energy and spin relaxation of a non-equilibrium electron gas in
gyrotropic systems can also drive an electric current. The current
is a result of relaxation of heated carriers, and hence its
magnitude and direction are independent of the polarization of
radiation. Several mechanisms related to the asymmetry of electron
relaxation are considered below.

\begin{figure}
\centerline{\epsfxsize=0.75\linewidth \epsfbox{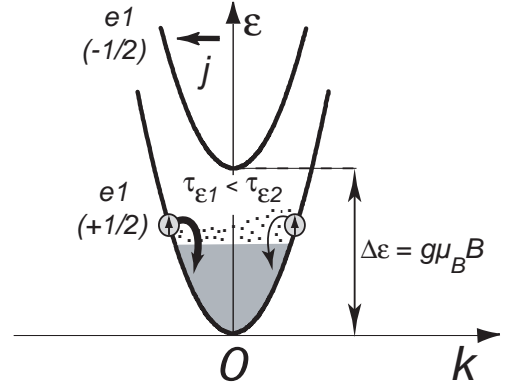}}
\caption{Microscopic origin of the electric current caused by
asymmetry of the energy relaxation in the presence of an in-plane
magnetic field. The spin subband $(+1/2)$ is preferably occupied
due to the Zeeman splitting. The $\bm{k}$-linear spin splitting is
neglected in the band structure because it is unimportant for this
mechanism.} \label{model2}
\end{figure}

\paragraph*{5.4.1. Asymmetry of electron energy relaxation.}
\label{1.2.1.}

Another mechanism which stems from spin-dependent asymmetric terms
in the electron-phonon interaction is the energy relaxation of hot
carriers~\cite{bulli}. The light absorption by free electrons
leads to an electron gas heating, i.e. to a non-equilibrium energy
distribution of electrons. Here we assume, for simplicity, that
the photoexcitation results in isotropic non-equilibrium
distribution of carriers. Due to asymmetry of electron-phonon
interaction discussed above, (see Eq.~(\ref{scattering}) and
Section~5.3.1.) hot electrons with opposite $\bm{k}$ have
different relaxation rates. This situation is sketched in
Fig.~\ref{model2} for a spin-up subband $(s=1/2)$, where two
arrows of different thicknesses denote non-equal relaxation rates.
As a result, an electric current is generated. Whether $-k$ or
$+k$ states relax preferentially, depends on the spin direction.
It is because the electron-phonon asymmetry is spin-dependent and
has the opposite sign in the other spin subband. Similarly to the
case described in the mechanism~5.3.1, the arrows in
Fig.~\ref{model2} need to be interchanged for the other spin
subband. For $B=0$ the currents in the spin-up and spin-down
subbands have opposite directions and cancel exactly. But as
described in Section 5.3.1 a pure spin current flows which
accumulates opposite spins at opposite edges of the sample. In the
presence of a magnetic field the currents moving in the opposite
directions do not cancel due to the non-equal population of the
spin subbands (see Fig.~\ref{model2}) and a net electric current
flows.

For the electron-phonon interaction given by
Eq.~(\ref{scattering}) one has
\begin{equation} \label{vdr}
V_{s_x \bm{k}', s_x \bm{k}} = \Xi_c \epsilon_{ii}\,  - \Xi_{cv}\
\xi  (k'_y + k_y)\ \epsilon_{xy}\ {\rm sign}\,s_x\:.
\end{equation}
Thus, the ratio of antisymmetric to symmetric parts of the
scattering probability rate, $W_{s_x {\bm k}', s_x {\bm k}}
\propto |V_{s_x \bm{k}', s_x \bm{k}}|^2$, is given by $W_{\rm
as}/W_{\rm s} \sim (\Xi_{cv} \xi \epsilon_{xy}/\Xi_c
\epsilon_{ii}) (k'_y + k_y)$. Since the antisymmetric component of
the electron distribution function decays within the momentum
relaxation time $\tau_p$, one can write for the photocurrent
\[
j_i \sim e N \frac{g \mu_B B_x}{\bar{\varepsilon}}
\]
\vspace{-0.6cm}
\[
\times \left\langle W_{\rm s}\ \frac{\Xi_{cv} \xi}{\Xi_c}\ \frac{
\epsilon_{xy} (k'_y + k_y)}{\epsilon_{ii}} \left[ \tau_p(k')
\frac{\hbar k'_i}{m^*} - \tau_p(k) \frac{\hbar k_i}{m^*} \right]
\right\rangle \:,
\]
where $N$ is the 2D electron density and the angle brackets mean
averaging over the electron energy distribution. While the average
for $j_y$ is zero,  the $x$ component of the photocurrent can be
estimated as
\begin{equation} \label{2}
j_x \sim \frac{e \tau_p}{\hbar}\ \frac{\Xi_{cv} \xi}{\Xi_c}\
\frac{g \mu_B B_x}{\bar{\varepsilon}}\ \eta I\:,
\end{equation}
where $\eta$ is the fraction of the energy flux absorbed in the QW
due to all possible indirect optical transitions. By deriving this
equation we took into account the balance of energy
$$
\sum_{{\bm k}' {\bm k}}[\varepsilon({\bm k}) - \varepsilon({\bm
k}')] W_{{\bm k}',{\bm k}} = \eta I \ ,
$$
where $\varepsilon({\bm k}) = \hbar^2 k^2/2 m^*$. An additional
contribution to the relaxation photocurrent comes if we neglect
the asymmetry of electron-phonon interaction by setting $\xi =0$
but, instead, take cubic-${\bm k}$ terms into account  in the
electron effective Hamiltonian.

Compared to the mechanisms~5.3, the main characteristic feature of
mechanism~5.4.1 is its independence of the in-plane
linear-polarization orientation, i.e. $S_2=S'_2=S_3=S'_3=0$. A
particular choice of $V_{s_x \bm{k}', s_x \bm{k}}$ in the form of
Eq.~(\ref{vdr}) leads to a photocurrent with $S'_1 = S_1$ or,
equivalently, $S^-_1 = 0$. By adding a spin-dependent invariant of
the type ${\cal I}_2$ to the right-hand part of Eq.~(\ref{vdr})
one can also obtain a nonzero value of $S^-_1$.

\paragraph*{5.4.2. Current due to spin-dependent asymmetry of spin relaxation
(spin-galvanic effect).} \label{1.2.2.}

This mechanism is based on the asymmetry of spin-flip relaxation
processes and represents in fact the spin-galvanic
effect~\cite{Nature02} where the current is linked to spin
polarization
\begin{equation}
j_i = Q_{ii'} (S_{i'} - S^{(0)}_{i'})\:. \label{Q}
\end{equation}
Here ${\bm S}$ is the average electron spin and ${\bm S}^{(0)}$ is
its equilibrium value, see Eq.~(\ref{spineq}).
In contrast to the majority of the mechanisms considered above
which do not contain $\bm
k$-linear terms, these are crucial here

In the previous  considerations the spin-galvanic
effect was described for a non-equilibrium spin polarization
achieved by optical orientation where ${\bm S}^{(0)}$ was
negligible~\cite{Nature02,review2003spin}. Here we discuss a more
general situation a non-zero ${\bm S}^{(0)}$ caused by the
Zeeman splitting in a magnetic field  is explicitly taken into account.
We show below that
in addition to
optical orientation with circularly polarized light, it opens  a new
possibility to achieve a non-equilibrium spin polarization and,
hence, an additional contribution to the photocurrent.

\begin{figure}
\centerline{\epsfxsize=0.75\linewidth \epsfbox{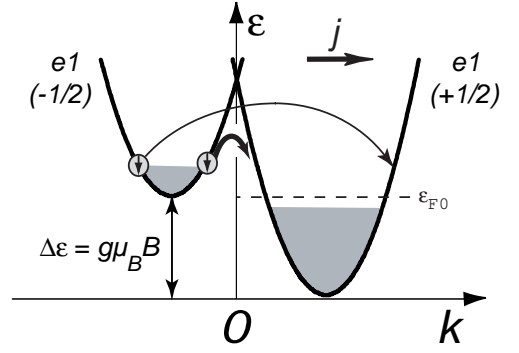}}
\caption{Microscopic origin of the electric current caused by
asymmetry of spin relaxation. Non-equilibrium spin is due to
photoinduced depolarization of electron spins. Asymmetry of spin
relaxation and, hence, an electric current is caused by
$\bm{k}$-linear spin splitting. } \label{model3}
\end{figure}

Fig.~\ref{model3} illustrates this mechanism. In equilibrium the electrons
preferably occupy the Zeeman split lower spin subband.
By optical excitation with light of any polarization a
non-equilibrium population as sketched in Fig.~\ref{model3} can be
achieved. This is a consequence of the fact that optical
transitions from the highly occupied subband dominate. These
optically excited electrons under energy relaxation return to both
subbands. Thus, a  non-equilibrium population of the spin subbands
appears. To return to equilibrium spin-flip transitions are
required. Since spin relaxation efficiently depends on initial and
final $k$-vectors, the presence of $k$-linear terms leads to an
asymmetry of spin relaxation (see bent arrows in
Fig.~\ref{model3}), and hence to current flow. This mechanism was
described in~\cite{Nature02}.

Following similar arguments as in Ref.~\cite{Nature02,PRB03sge}
one can estimate the spin-galvanic contribution to the
polarization-independent magneto-induced photocurrent as
\begin{equation}
j \sim e \tau_p \frac{\beta}{\hbar} \frac{g \mu_B
B}{\bar{\varepsilon}} \frac{\eta I}{\hbar \omega}\: \zeta\:.
\end{equation}
Here $\zeta$ is a factor describing the electron spin
depolarization due to photoexcitation followed by the energy
relaxation. It can be estimated as $\zeta \sim
\tau_{\varepsilon}/\tau_{s}$, where $\tau_{\varepsilon}$ is
electron energy relaxation time governed mainly by
electron-electron collisions, and $\tau_{s}$ is the spin
relaxation time. Assuming $\tau_{\varepsilon} \sim 10^{-13}\,s$
and $\tau_{s} \sim 10^{-10}\,s$ at room temperature, the factor
$\zeta$ is estimated as $10^{-3}$.

\subsection*{ 5.5. Spin-independent mechanisms of magneto-induced
photocurrent} \label{diamagnetic}

The last group of mechanisms is based on  a magnetic field induced
shift of the energy dispersion in the ${\bm k}$-space in
gyrotropic materials. This mechanism was investigated
theoretically and observed experimentally for direct inter-band
transitions~\cite{moscow,kucher} and proved to be efficient. To
obtain such a current for indirect optical transitions one should
take into account effects of the second order like
non-parabolicity or transitions via virtual states in the other
bands. Our estimations show that these processes are less
efficient compared to mechanisms~5.3 and 5.4. However, to be
complete, we consider below possible contributions of the
diamagnetic shift to the current at the Drude absorption of
radiation.

\paragraph*{5.5.1. Spin-independent asymmetry of indirect transitions
with intermediate states in the same subband.} \label{2.1.1.}

The experiments on the MPGE under direct optical transitions
observed in asymmetric QW structures are interpreted in terms of
the asymmetric spin-independent electron energy dispersion,
$\varepsilon({\bm k}, {\bm B}) \neq \varepsilon(-{\bm k}, {\bm
B})$, analyzed by Gorbatsevich et al. \cite{gorbats}, see also
\cite{kibis,spivak}. The simplest contribution to the electron
effective Hamiltonian representing such kind of asymmetric
dispersion is the diamagnetic term ${\cal H}^{\rm (dia)}_{\rm
SIA}$ in Eq.~(\ref{diamaH}). In asymmetric QWs, $\bar{z}_{\nu}$
are nonzero and the subband dispersion is given by parabolas with
their minima (or maxima in case of the valence band) shifted from
the origin $k_x=k_y=0$ by a value proportional to the in-plane
magnetic field.

For  indirect optical transitions these linear-${\bm k}$ terms do
not lead, in the first order, to a photocurrent. To obtain the
current one needs to take into account the non-parabolic
diamagnetic term
\begin{equation} \label{dia3}
{\cal H}^{\rm (dia,3)}_{\rm SIA} = {\cal F}_{\rm SIA} (B_x k_y -
B_y k_x) k^2\:.
\end{equation}
The non-parabolicity parameter can be estimated by ${\cal F}_{\rm
SIA} \sim (\hbar^2/m^*E_g)\: \tilde{\alpha}_{\rm SIA}$. By analogy
with the SIA diamagnetic term we can introduce the BIA diamagnetic
term ${\cal H}^{\rm (dia,3)}_{\rm BIA} = {\cal F}_{\rm BIA} (B_x
k_x - B_y k_y)k^2$. It is most likely that, in realistic QWs, the
coefficient ${\cal F}_{\rm BIA}$ is small as compared to ${\cal
F}_{\rm SIA}$.

\paragraph*{5.5.2. Spin-independent asymmetry of indirect transitions via
other bands and subbands.} \label{2.1.2.}

One can show, that even the linear-${\bm k}$ diamagnetic terms can
contribute to the photocurrent under indirect intra-subband
optical transitions if the indirect transition involves
intermediate states in other bands (or subbands) different from
the conduction subband $e1$. Under normal incidence of the light,
a reasonable choice could be a combination of direct intra-band
optical transitions with the piezoelectric electron-phonon
interaction, for the first process, and inter-band virtual optical
transitions as well as interband deformation-potential
electron-phonon interaction, for the second process. An asymmetry
of the indirect photoexcitation is obtained as a result of the
interference between two indirect processes with the intermediate
state in the same subband and elsewhere. Moreover, the diamagnetic
dispersion asymmetry of the initial and intermediate bands should
be taken into account in the energy denominator of the compound
two-quantum matrix element for the transitions via other bands.

\paragraph*{5.5.3.  Spin-independent asymmetry of electron energy
relaxation.} \label{2.2}

Similarly to the spin-dependent mechanism 5.4.1, the diamagnetic
cubic-${\bm k}$ term, see Eq.~(\ref{dia3}), can be responsible for
the relaxational photocurrent. This relaxation mechanism is
unlikely to give an essential contribution to the MPGE.

To summarize this group of mechanisms we note that, as in the case
of spin-dependent mechanisms, the mechanisms 5.5.1 and 5.5.2 allow
a pronounced dependence of the photocurrent on the orientation of
the in-plane light polarization whereas the relaxation mechanism
5.5.3 is independent of the polarization state.

%%%%%%%%%%%%%%%%%%%%%%%%%%%%%%%%%%%%%%%%%%%%%%%%%%%%%%%%%%%%%%%%

\section{Discussion}

In all investigated QW structures, an illumination with terahertz
radiation in the presence of an in-plane magnetic field results in
a photocurrent in full agreement with the phenomenological theory
described  by Eqs.~(\ref{phen}). The microscopic treatment
presented in Section~\ref{micmodel} shows that  two classes of
mechanisms dominate the magneto-gyrotropic effects. The current
may be  induced either by an asymmetry of optical excitation
and/or by an asymmetry of relaxation. Though in all cases the
absorption is mainly independent of the light polarization, the
photocurrent depends on polarization for the first class of the
mechanisms (see Section~5.3) but is independent of the direction
of linear light polarization for the second class (see
Section~5.4). Thus the polarization dependence of the
magneto-gyrotropic photocurrent signals allows us to distinguish
between the above two classes. The asymmetry of photoexcitation
may contribute to all terms in Eqs.~(\ref{phen}). Therefore, such
photocurrent contributions should exhibit a characteristic
polarization dependence given, for linearly polarized light, by
the second and third terms in Eqs.~(\ref{phen}) with the
coefficients $S_2, S'_2, S_3, S'_3$. In contrast, the asymmetry of
relaxation processes (see Section~5.4) contributes only to the
coefficients $S_1,S'_1, S_4, S'_4$.

The experimental data obtained on the samples A1 to A4 suggest
that in these QW structures relaxation mechanisms, presented in
Section 5.4, dominate. Indeed only current contributions described
by the first and last terms in Eqs.~(\ref{phen})  are detectable,
whereas the second and third term contributions are vanishingly
small. These samples are denoted as type I below. The results
obtained for type I samples are valid in the wide temperature
range from 4.2 K up to room temperature. The transverse
photocurrent observed in the direction normal to the magnetic
field $\bm{B}$ applied along $\langle 110 \rangle$ is independent
of the light polarization. It corresponds to the first term in
Eqs.~(\ref{phen}). Hence, this current is caused by the Drude
absorption-induced electron gas heating  followed by energy
relaxation (mechanism 5.4.1) and/or spin relaxation (mechanism
5.4.2). The analysis (see Section 5.4) shows that in the absence
of the magnetic field electron gas heating in gyrotropic QWs is
accompanied by a pure spin flow. The longitudinal photocurrent
component parallel to $\bm{B}$, which appears under excitation
with circularly polarized radiation only, arises due to spin
relaxation of optically oriented carriers (spin-galvanic
effect~\cite{Nature02,review2003spin}).

In contrast to the samples of type I, the experimental results
obtained on the sample A5 (in the following denoted as type II)
has characteristic polarization dependencies corresponding to the
second ($S_2, S'_2$) and third ($S_3, S'_3$) terms in
Eqs.~(\ref{phen}). The photocurrent exhibits a pronounced
dependence on the azimuthal  angle $\alpha$ of the linear
polarization, but it is equal for the right and left circular
polarized light. This experimental finding proves that the main
mechanism for current generation in type II sample  is the
asymmetry of photoexcitation considered in Section~5.3.

The question concerning the difference of type I and type II
samples remains open.
While experimentally the two classes of the mechanisms are clearly
observed, it is not clear yet what determines large difference
between the relevant $S$-coefficients.
Not much difference is expected between the type I and II samples
regarding  the strength and asymmetry of electron-phonon
interaction. The samples only differ in the type of doping and the
electron mobility. The influence of impurity potentials (density,
position, scattering mechanisms etc.)  on microscopic level needs
yet to be explored. In addition, the doping level of the type I
samples is significantly lower and the mobility is higher than
those in the type II samples. This can also affect the interplay
between the excitation and relaxation mechanisms.

Finally we note, that under steady-state optical excitation, the
contributions of the relaxation and photoexcitation mechanisms to
magneto-induced photogalvanic effects are superimposed. However,
they can be separated experimentally in time-resolved
measurements. Indeed, under the ultra-short pulsed photoexcitation
the current should decay, for the mechanisms considered above,
within the energy ($\tau_{\varepsilon}$), spin ($\tau_s$) and
momentum ($\tau_p$) relaxation times times.

%%%%%%%%%%%%%%%%%%%%%%%%%%%%%%%%%%%%%%%%%%%%%%%%%%%%%%%%%%%%%%%%%

\section{Summary}

%%%%%%%%%*****
We have studied  photocurrents  in $n$-doped zinc-blende based
(001)-grown QWs generated by the Drude absorption of normally
incident terahertz radiation in the presence of an in-plane
magnetic field. The results agree with the phenomenological
description based on the symmetry.
Both experiment and theoretical analysis show that
there are a variety of routes  to generate   spin polarized
currents. As we used both magnetic fields and gyrotropic
mechanisms we coined the notation ''magneto-gyrotropic
photogalvanic effects'' for this class of phenomena.

\subsection*{Acknowledgements} The high quality InAs quantum wells
were kindly provided by J.~De~Boeck and G.~Borghs from IMEC
Belgium.  We thank L.E.~Golub for helpful discussion. This work
was supported by the DFG, the RFBR, the INTAS, programs of the
RAS, and Foundation ``Dynasty'' - ICFPM.

%%%%%%%%%%%%%%%%%%%%%%%%%%%%%%%%%%%%%%%%%%%%%%%%%%%%%%%%%%%%%%%%%%

\begin{widetext}

\section{Appendices}

\subsection*{8.1. Appendix A. Point Groups T$_d$ and D$_{2d}$}

In the T$_d$\,-\,class bulk crystals the MPGE linear in the
magnetic field ${\bm B}$ can be phenomenologically presented
as~\cite{ferro,sovrem}
\begin{equation}\label{phend}
j_x = 2 S_1 \left( |e_y|^2 - |e_z|^2 \right) B_x I + S_2 \left[
\left( e_z e_x^* + e_x e_z^* \right) B_z  - \left( e_x e_y^* + e_y
e_x^* \right) B_y \right] I - S_4 \left[ {\rm i} \left( {\bm e}
\times {\bm e}^* \right)_y B_z + {\rm i} \left( {\bm e} \times
{\bm e}^* \right)_z B_y \right] I\:,
\end{equation}
and similar expressions for $j_y$ and $j_z$, where $x \,
\|\,[100],\, y \, \|\, [010],\, z \, \|\,[001]$. Note that here
the notation of the coefficients is chosen as to be in accordance
with the phenomenological equations~(\ref{phena}). Under
photoexcitation along the [001] axis, $e_z = 0$ and, in the
presence of an external magnetic field ${\bm B} \perp [001]$, one
has
\begin{eqnarray} \label{phend1}
j_x &=& S_1 [1 - ( |e_x|^2 - |e_y|^2)] B_x I - B_y I \left[ S_2
\left( e_x e_y^* + e_y e_x^* \right) + S_4 P_{\rm circ} \right]
\:,
\\ j_y &=& - S_1 [1 + ( |e_x|^2 - |e_y|^2)] B_y I + B_x I \left[ S_2
\left( e_x e_y^* + e_y e_x^* \right) - S_4 P_{\rm circ} \right]
\:. \nonumber
\end{eqnarray}
In the axes $x^{\prime}\, \|\, [1 \bar{1} 0],\, y^{\prime} \, \|\,
[110],\, z \, \|\, [001]$, Eqs.~(\ref{phend1}) assume the form
\begin{eqnarray} \label{phen3}
j_{x^{\prime}} &=& S_1 \left[ B_{y^{\prime}} - \left(
e_{x^{\prime}} e^*_{y^{\prime}} + e_{y^{\prime}} e^*_{x^{\prime}}
\right) B_{x^{\prime}} \right] I + S_2 \left( |e_{x^{\prime}}|^2 -
|e_{y^{\prime}}|^2 \right) B_{y^{\prime}} I + S_4 P_{\rm circ}
B_{x^{\prime}}I\:, \\  j_{y^{\prime}} &=& S_1 \left[
B_{x^{\prime}} - \left( e_{x^{\prime}} e^*_{y^{\prime}} +
e_{y^{\prime}} e^*_{x^{\prime}} \right) B_{y^{\prime}} \right]I -
S_2 \left( |e_{x^{\prime}}|^2 - |e_{y^{\prime}}|^2 \right)
B_{x^{\prime}}I - S_4 P_{\rm circ} B_{y^{\prime}}I\:. \nonumber
\end{eqnarray}

Equations (\ref{phend1},\ref{phen3}) are consistent with
Eqs.~(\ref{phen},\ref{phena}) describing the magneto-induced
photo\-currents in the $C_{2v}$-symmetry systems and can be
obtained from Eqs. (\ref{phen},\ref{phena}) by setting $S_1' = S_1
= - S_3 = - S'_3$, $S'_2 = - S_2$ or, equivalently, $S^-_1 = S^+_2
= S^-_3 = S^+_4 =0$ and $S^+_1 = - S^+_3 = S_1, S^-_2 = S_2, S^-_4
= S_4$.

One can show that the phenomenological equations for the D$_{2d}$
symmetry are obtained from Eqs.~(\ref{phen},\ref{phena}) if we set
$S_1' = S_1$, $S_3 = S'_3$, $S'_2 = - S_2$, $S'_4 = - S_4$. The
only difference with Eqs. (\ref{phend1}, \ref{phen3}) is that
$S_1$ and $S_3$ are now linearly independent.

%%%%%%%%%%%%%%%%%%%%%%%%%%%%%%%%%%%%%%%%%%%%%%%%%%%%%%%%%%%%%%%%%%

\subsection*{8.2. Appendix B. Point Group C$_{\infty v}$}

For a system of the C$_{\infty v}$ symmetry, one has
\nix{\begin{eqnarray} \label{phen1}
j_{x^{\prime}} &=& S_1 I B_{y^{\prime}} + S_2 I \left[ \left(
|e_{x^{\prime}}|^2 - |e_{y^{\prime}}|^2 \right) B_{y^{\prime}} -
\left( e_{x^{\prime}} e^*_{y^{\prime}} + e_{y^{\prime}}
e^*_{x^{\prime}} \right) B_{x^{\prime}} \right] + S_4 I P_{\rm
circ} B_{x^{\prime}}\:, \\ j_{y^{\prime}} &=& - S_1 I
B_{x^{\prime}} + S_2 I \left[ \left( |e_{x^{\prime}}|^2 -
|e_{y^{\prime}}|^2 \right) B_{x^{\prime}} + \left( e_{x^{\prime}}
e^*_{y^{\prime}} + e_{y^{\prime}} e^*_{x^{\prime}} \right)
B_{y^{\prime}} \right] + S_4 I P_{\rm circ} B_{y^{\prime}}\:.
\nonumber
\end{eqnarray}
}
\begin{equation} \label{phenb}
j_x = S_1 B_y I + S_2 \left[ \left( |e_x|^2 - |e_y|^2 \right) B_y
 - \left( e_x e_y^* + e_y e_x^*
\right) B_x \right] I + S_4
B_x I P_{\rm circ}\:,
\end{equation}
$$
j_y = - S_1 B_x I + S_2 \left[\left( |e_x|^2 - |e_y|^2 \right) B_x  + \left( e_x e_y^* + e_y e_x^*
\right) B_y  \right] I + S_4
B_y I P_{\rm circ}\:.
$$
where the  form of the equation is independent of the orientation
of Cartesian coordinates $(x,y)$ in a plane normal to the
C$_{\infty}$-axis. A comparison to Eqs.~(\ref{phen}) for C$_{2v}$
symmetry shows that the form of these equations is identical
besides the coefficients $S_i$. In this case we have $S'_1 = -
S_1, S_2 = S'_2 = - S_3 = S'_3$, $S'_4 = S_4$.\\

\end{widetext}

\end{document}